\documentclass[12pt,preprint]{aastex} 
\newcommand{\iras}{IRAS~18317$-$0757}
\newcommand{\ceo}{C$^{18}$O}
\newcommand{\nceo}{$N({\rm C}^{18}{\rm O})$}
\newcommand{\ammonia}{NH$_3$} 
\newcommand{\lsun}{$L_\odot$}
\newcommand{\msun}{$M_\odot$}
\newcommand{\uchii}{UC\ion{H}{2}}
\shortauthors{Hunter et al.}
\shorttitle{The Protocluster \iras}
\begin{document}
\title{\iras: A Cluster of Embedded Massive Stars and Protostars} 
\author{T.R. Hunter, Q. Zhang, T.K. Sridharan}
\affil{Harvard-Smithsonian Center for Astrophysics, MS-78, 60 Garden St., 
Cambridge, MA 02138} 
\email{thunter@cfa.harvard.edu}


\begin{abstract}

We present high resolution, multiwavelength continuum and molecular
line images of the massive star-forming region \iras.  The global
infrared through millimeter spectral energy distribution can be
approximated by a two temperature model (25K and 63K) with a total
luminosity of approximately log($L/L_\odot$)~=~5.2.  Previous
submillimeter imaging resolved this region into a cluster of five dust
cores, one of which is associated with the ultracompact \ion{H}{2}
region G23.955+0.150, and another with a water maser.  In our new
2.7mm continuum image obtained with BIMA, only the \uchii\ region is
detected, with total flux and morphology in good agreement with the
free-free emission in the VLA centimeterwave maps. For the other four
objects, the non-detections at 2.7mm and in the MSX mid-infrared bands
are consistent with cool dust emission with a temperature of 13-40K
and a luminosity of 1000-40000 \lsun.  By combining single-dish and
interferometric data, we have identified over two dozen virialized
\ceo\ cores in this region which contain $\approx 40\%$ of the total
molecular gas mass present.  While the overall extent of the \ceo\ and
dust emission is similar, their emission peaks do not correlate well
in detail.  At least 11 of the 123 infrared stars identified by 2MASS
in this region are likely to be associated with the star-forming
cluster. Two of these objects (both associated with \uchii) were
previously identified as O stars via infrared spectroscopy.  Most of
the rest of the reddened stars have no obvious correlation with the
\ceo\ cores or the dust continuum sources.  In summary, our
observations indicate that considerable fragmentation of the molecular
cloud has taken place during the time required for the \uchii\ region
to form and for the O stars to become detectable at infrared
wavelengths.  Additional star formation appears to be ongoing on the
periphery of the central region where up to four B-type (proto)stars
have formed amongst a substantial number of \ceo\ cores.

\end{abstract}


\section{Introduction}

The formation mechanism of massive stars is a topic of active
research.  Because massive star formation regions typically lie at
distances of several kiloparsecs, the identification of high mass
protostars requires both good sensitivity and high angular resolution.
As a consequence of their presumed youth, ultracompact HII regions
(\uchii\ regions) provide a good tracer of current massive star
formation \citep{Wood89} and may be expected to be accompanied by
protostars in earlier evolutionary stages.  Indeed, recent
high-resolution millimeterwave images of \uchii\ regions have revealed
the high-mass equivalent of ``Class~0'' protostars.  Examples include
the young stellar object IRAS~23385+6053 \citep{Molinari98}, the
compact methyl cyanide core near the G31.41+0.31 \uchii\
\citep{Cesaroni94}, the proto-B-star G34.24+0.14MM \citep{Hunter98},
the G9.62+0.19-F hot core \citep{Testi00}, and the protocluster
G24.78+0.08 \citep{Furuya02}.  To identify these deeply-embedded
objects requires an optically thin tracer in order to probe through
the large extinction toward the giant molecular cloud cores that
harbor them.  Submillimeter continuum emission from cool dust is a
good tracer of protostars because it remains optically thin at high
column densities ($N_{\rm H} \lesssim 10^{25} {\rm cm}^{-2}$)
\citep{Mezger94}.  Similarly, spectral line emission from \ceo\ is a
good optically thin tracer that can reveal areas of high molecular gas
column density.

Our target in this study, \iras, is a luminous infrared source (log
L$_{\rm FIR} = 5.2$) at a kinematic distance of 4.9 kpc ($v_{\rm LSR}$
= 80 km~s$^{-1}$).  Based on its IRAS colors, it has been identified
as a massive protostellar candidate \citep{Chan96}.  Previous
single-dish radio frequency studies of this region have revealed
hydrogen recombination line emission \citep{Kim01,Lockman89,Wink83}
and water maser emission \citep{Genzel77,Churchwell90}.  The
centimeterwave continuum emission shows both extended components (up
to $13'$) \citep{Kim01,Becker94} and a \uchii\ region \citep{Wood89}.
The region has been detected in various dense gas tracers including
the NH$_3$(1,1), (2,2) and (3,3) transitions \citep{Churchwell90} and
CS(7-6) \citep{Plume92}, though it was not detected in a methyl
cyanide search \citep{Pankonin01}, nor in a 6~GHz hydroxl maser search
\citep{Baudry97}, nor in two 6.7~GHz methanol maser searches
\citep{Szymczak00,Walsh97}. The CO(1-0) line shows a complex profile
which has prevented the identification of high-velocity outflow
emission in large-beam ($1'$) surveys \citep{Shepherd96}.

Also known as AFGL2194, compact infrared emission was first detected
from the ground in the $K$, $L$, and $M$ bands by \citet{Moorwood81} and
later by \citet{Chini87}.  Airborne observations of far-infrared
continuum and fine-structure lines (S, O, N, and Ne) yield an electron
density of 3500~cm$^{-3}$ for the \uchii\ region and indicate a
stellar type of O9 to early B \citep{Simpson95}.  Complete infrared
spectra ($2.4-195\mu$m) have been recorded by the ISO SWS and LWS
spectrometers \citep{Peeters02}. At higher angular resolution, the
region has been independently observed as part of two submillimeter
continuum imaging surveys. In both cases, the emission is resolved
into several components \citep{Hunter00,Mueller02}.  Recent
near-infrared imaging and spectroscopy has revealed the presence of a
small cluster of stars associated with the \uchii\ region, including
an O7 star (with \ion{N}{3} emission and \ion{He}{2} absorption) whose
ionizing flux can account for all of the compact radio continuum
emission \citep{Hanson02}.  These developments have prompted the
higher angular resolution millimeterwave observations which are
presented in this paper in hopes of understanding this active site of
massive star formation.

\section{Observations}

With the Berkeley-Illinois-Maryland-Association (BIMA) Millimeter
Array \citep{Welch96}, \iras\ was simultaneously observed in 110~GHz
continuum and \ceo\ (1-0).  The continuum bandwidth was 600~MHz.  The
spectral resolution for the line data was 0.2~MHz (0.53~km~s$^{-1}$).
The phase gain calibrator was the quasar 1741$-$038.  The bandpass
calibrator was 3C273.  The absolute flux calibration is based on 3C273
and Uranus.  A single track in B-configuration was obtained on 1998
October 10, and in C-configuration on 1999 February 5.  The
synthesized beam for the combined data imaged with robust weighting is
$4.4''$ by $2.3''$ at a position angle of $-2^\circ$.


To recover the missing flux from extended structures in the
interferometer spectral line data, a single-dish map of \ceo\ (1-0)
was recorded at the NRAO\footnote{The National Radio Astronomy
Observatory is a facility of the National Science Foundation operated
under cooperative agreement by Associated Universities, Inc.} 12 Meter
telescope on 16 June 2000.  A regular grid of 7x7 points was observed,
with a spacing of $25''$ providing full sampling of the telescope's
$58''$ beam.  The system temperature was 200K and the on-source
integration time was 2.8 minutes per point.  The data were combined as
zero-spacing information with the BIMA data in the MIRIAD
(Multichannel Image Reconstruction, Image Analysis and Display)
software package.  The resulting beamsize in the final datacube (with
natural weighting and UV tapering applied) is $10.3'' \times 7.1''$ at
a position angle of $-18^\circ$.

To complement the millimeter data, infrared images and point source
information for this region in $J$, $H$ and $K$ bands were obtained from the
2 Micron All Sky Survey (2MASS) \citep{Cutri2003} and in the
mid-infrared bands from the Midcourse Space Experiment (MSX) archives
\citep{Egan99} and HIRES-processed IRAS data \citep{Hunter97}.  Radio 
continuum images were also retrieved from the VLA galactic plane survey 
of \citet{Becker94}.

\section{Results} 

\subsection{Millimeter and radio continuum}

\begin{figure}
\figcaption[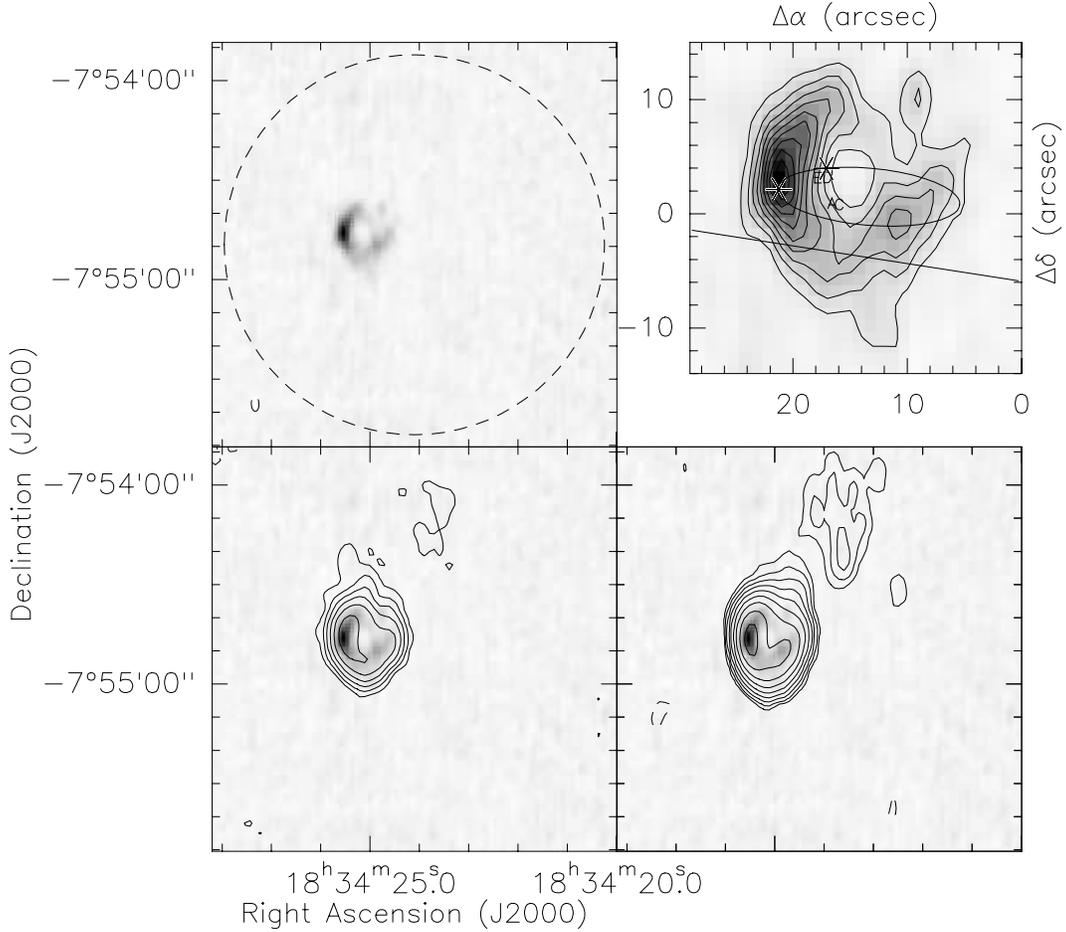]{In all panels, the grayscale is the
2.7 mm continuum imaged with a $4.4''$ by $2.3''$ synthesized
beam. Contours: upper right (2.7 mm): 3.6 mJy/beam $\times$ (4 to 36 by 4);
lower left (6 cm): 1.27 mJy/beam $\times$ (-3, 3, 6, 12, 24, 48, 96); lower
right (20 cm): 0.35 mJy/beam $\times$ (-3, 3, 6, 12, 24, 48, 96, 192, 384,
768).  Both centimeter maps were generated from data originally 
published by \citet{Becker94}.  In the
upper left panel, the dashed circle marks the BIMA primary beam at
half-maximum response.  In the upper right panel, the letters A, C, D
and E correspond to the fitted position of the peak emission in the
corresponding MSX band ($8.3\mu$m, $12.1\mu$m, $14.7\mu$m, $21.4\mu$m)
all of which are contained in the IRAS PSC error ellipse.  The six-pointed
stars indicate the positions of 2MASS star 86 (nearest to the 2.7mm peak) and
star~92. They are classified as O8.5 and O7, respectively
\citep{Hanson02}. The line marks the axis used to produce
Figure~\ref{stripchan34}.
\label{bimacont}
}
\epsscale{0.85}
\plotone{f1.eps}
\end{figure}

The 2.7mm continuum image from our BIMA observations is shown in
grayscale in Figure~\ref{bimacont}, along with overlays of the 6cm
and 20cm contours from the VLA galactic plane survey images
\citep{Becker94}.  At all three wavelengths, the source structure
consists of a bright rim on one edge of a partially-complete shell.
Both the IRAS point source and the MSX point source positions lie
close to the middle of the shell region.  The bright, compact
component was identifed as an irregular/multiple-peaked \uchii\ region
by \citet{Wood89}.  At the two longest radio wavelengths, additional
faint emission extends to the northwest.

\subsection{Submillimeter continuum}

\begin{figure}
\figcaption[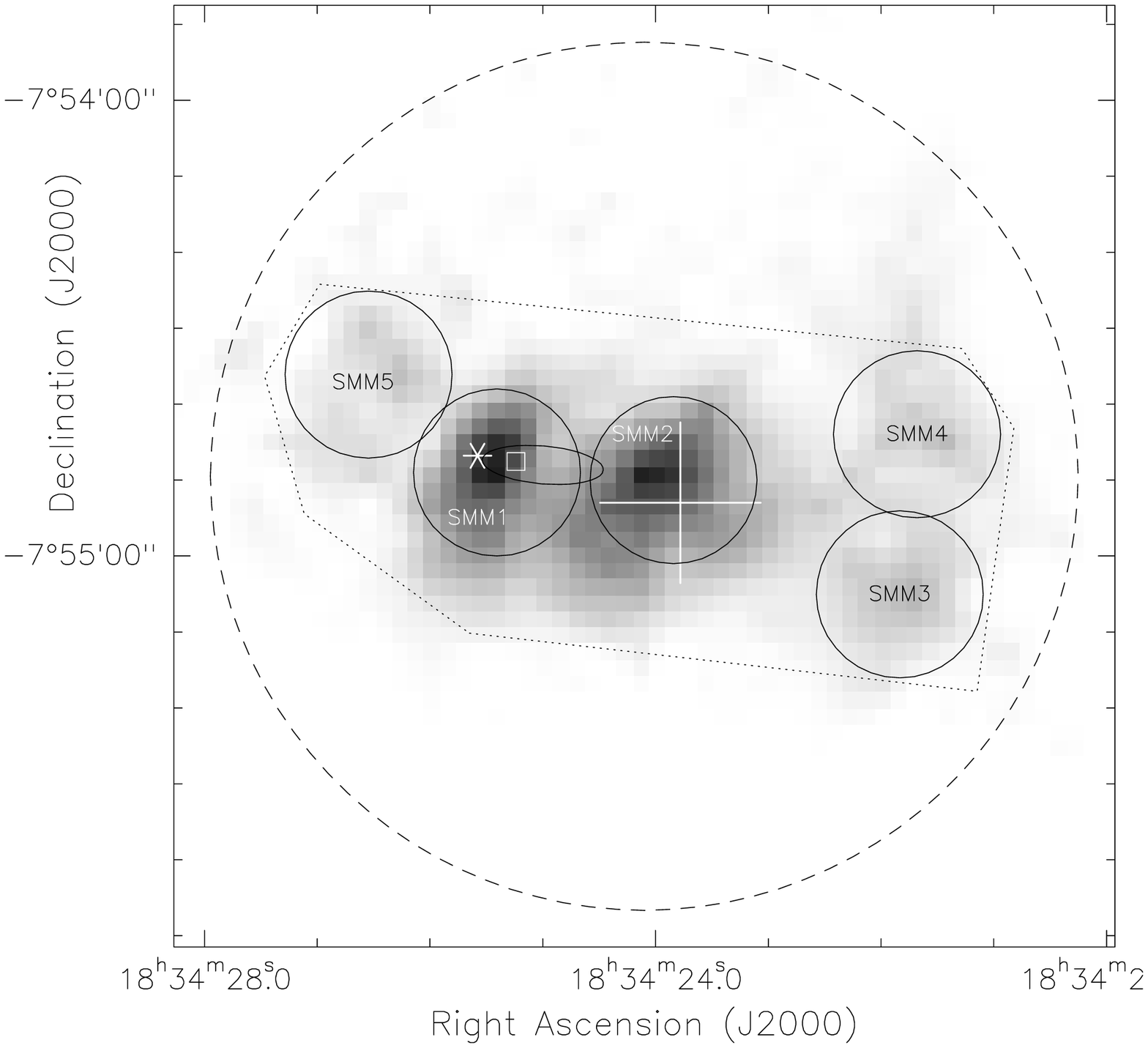]{350 micron continuum image of
\iras\ observed with an $11''$ beam \citep{Hunter00}.  The \uchii\
position is marked by the star symbol.  The cross marks the water
maser position uncertainty \citep{Genzel77}. The square contains the
fitted peak of the single point source seen in all four MSX bands,
which is contained by the IRAS error ellipse. The dashed circle marks
the BIMA primary beamsize, for reference to Figures~\ref{msx},
\ref{c18omom0} and \ref{2massk}. The circles and dotted line 
defines the point sources and extended emission region listed 
in Table~\ref{sharctable}.
\label{sharc}
}
\epsscale{1.0}
\plotone{f2.eps}
\end{figure}

The 350 micron continuum image from the survey of \citet{Hunter00} is
shown in grayscale in Figure~\ref{sharc}. For comparison, the position
of the IRAS and MSX point sources are indicated along with the
single-dish water maser position.  Five independent submillimeter
sources can be identified and their coordinates and flux densities are
given in Table~\ref{sharctable}.  The two dominant sources are SMM1
and SMM2. The peak of SMM1 coincides with the \uchii\ position.  SMM2
lies $22''$ to the west, and likely coincides with the water maser
emission, whose position is uncertain to $\pm10''$ (no interferometric
observations exist).  The water maser is apparently quite variable
over time: 60~Jy in 1976 \citep{Genzel77} to 0.7~Jy in 1989
\citep{Churchwell90} to undetected in 2002 (H. Beuther 2003, private
communication).  The three other sources have no known counterpart at
other wavelengths.

\subsection{Mid-infrared continuum}

\begin{figure}
\figcaption[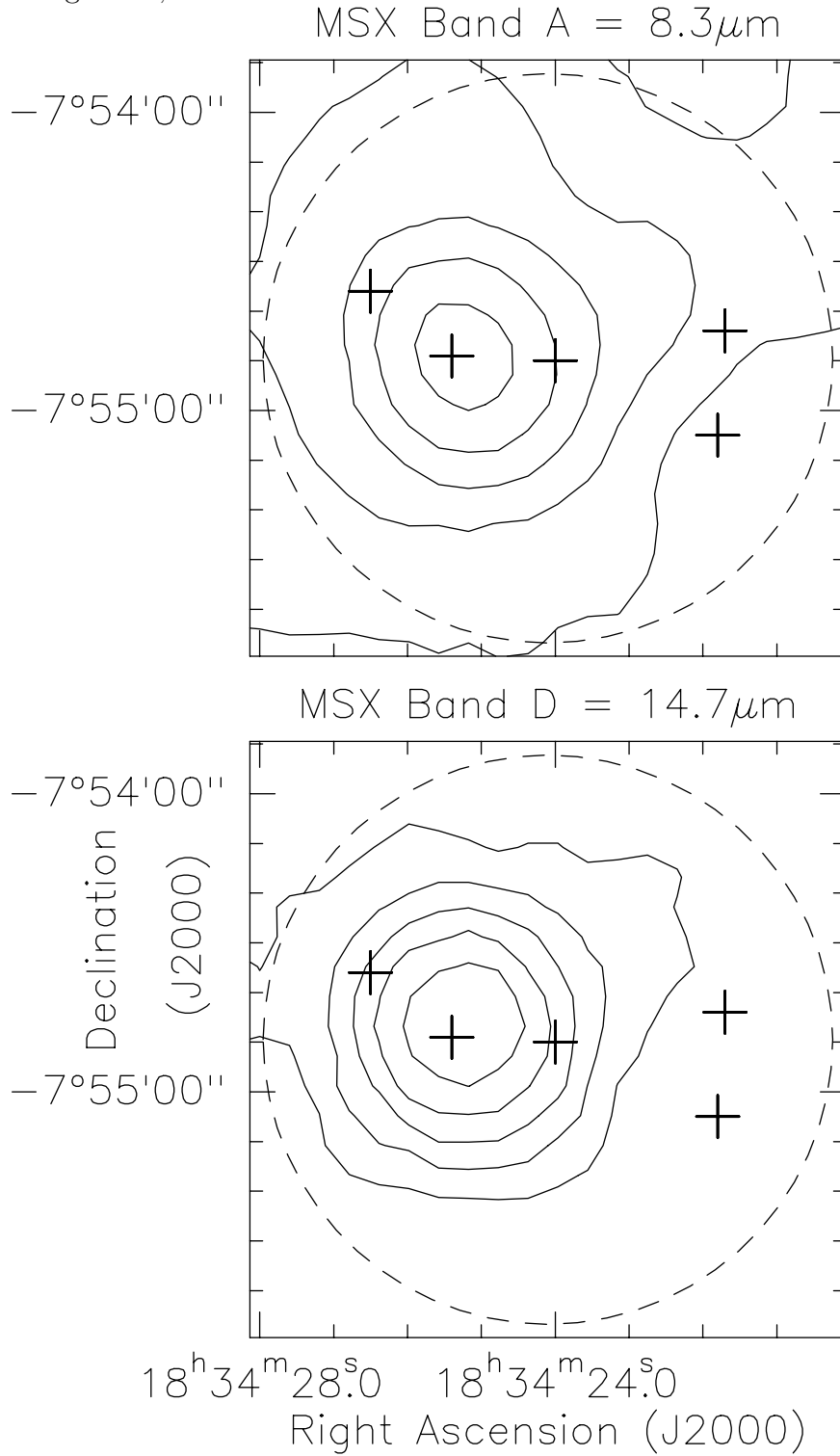]{MSX images of \iras. Contour levels are 0.005,
0.01, 0.02, 0.04, 0.08, 0.16, 0.32 and 0.64 erg cm$^{-2}$ s$^{-1}$
steradian$^{-1}$.  The crosses mark the position of the submillimeter
sources SMM1-5 (Table~\ref{sharctable}).  The dashed circle marks the
BIMA primary beamsize, for reference to Figures~\ref{sharc},
\ref{c18omom0} and \ref{2massk}.
\label{msx}
}
\epsscale{0.70}
\plotone{f3.eps}
\end{figure}

\begin{figure}
\figcaption[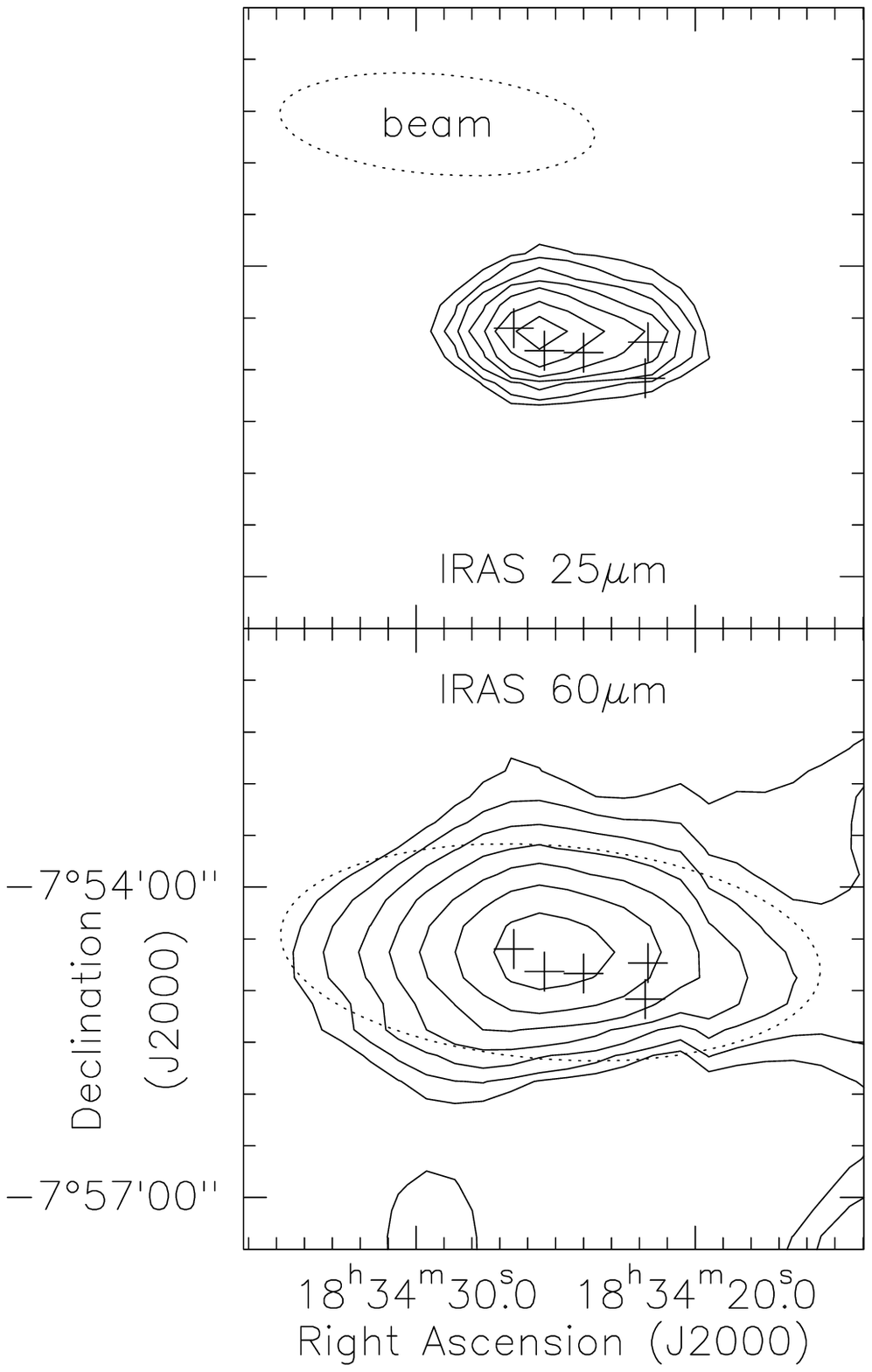]{IRAS HIRES-processed images of
\iras. Gaussian fits to the HIRES restoring beams are denoted by
dotted ellipses.  The crosses mark the position of the submillimeter
sources SMM1-5 (Table~\ref{sharctable}).  The contour levels are in
MegaJy steradian$^{-1}$: 
$25\mu$m (190, 381, 762, 1523, 3047, 6093, 12186), 
$60\mu$m (258, 516, 1032, 2064, 4128, 8257, 16513). 
\label{hires}
}
\epsscale{0.75}
\plotone{f4.eps}
\end{figure}

Each of the MSX images of this field show that the mid-infrared
emission is dominated by the \uchii\ region associated with SMM1.
Contour plots of two of the bands (8.3 and $14.7\mu$m) are shown in
Figure~\ref{msx}.  The positions of the five submillimeter sources are
indicated in both panels.  The fitted positions of a two-dimensional
Gaussian model in each of the four MSX bands all agree to within $2''$
and fall within the IRAS PSC error ellipse.  Of the five submillimeter
sources, the peak in each band lies closest to SMM1.  HIRES-processed
images provide additional high resolution information from the IRAS
data \citep{Aumann90}.  The 20-iteration contour maps at 25 and
$60\mu$m are shown in Figure~\ref{hires}, again with the five
submillimeter sources marked.  The emission remains essentially
unresolved in each band, though there is some hint that the two
westernmost submillimeter sources (SMM3 and 4) are detected in the
contour extensions at 25 and 60$\mu$m.

\subsection{Spectral Energy Distributions}
\label{sedsection}
\begin{figure}
\figcaption[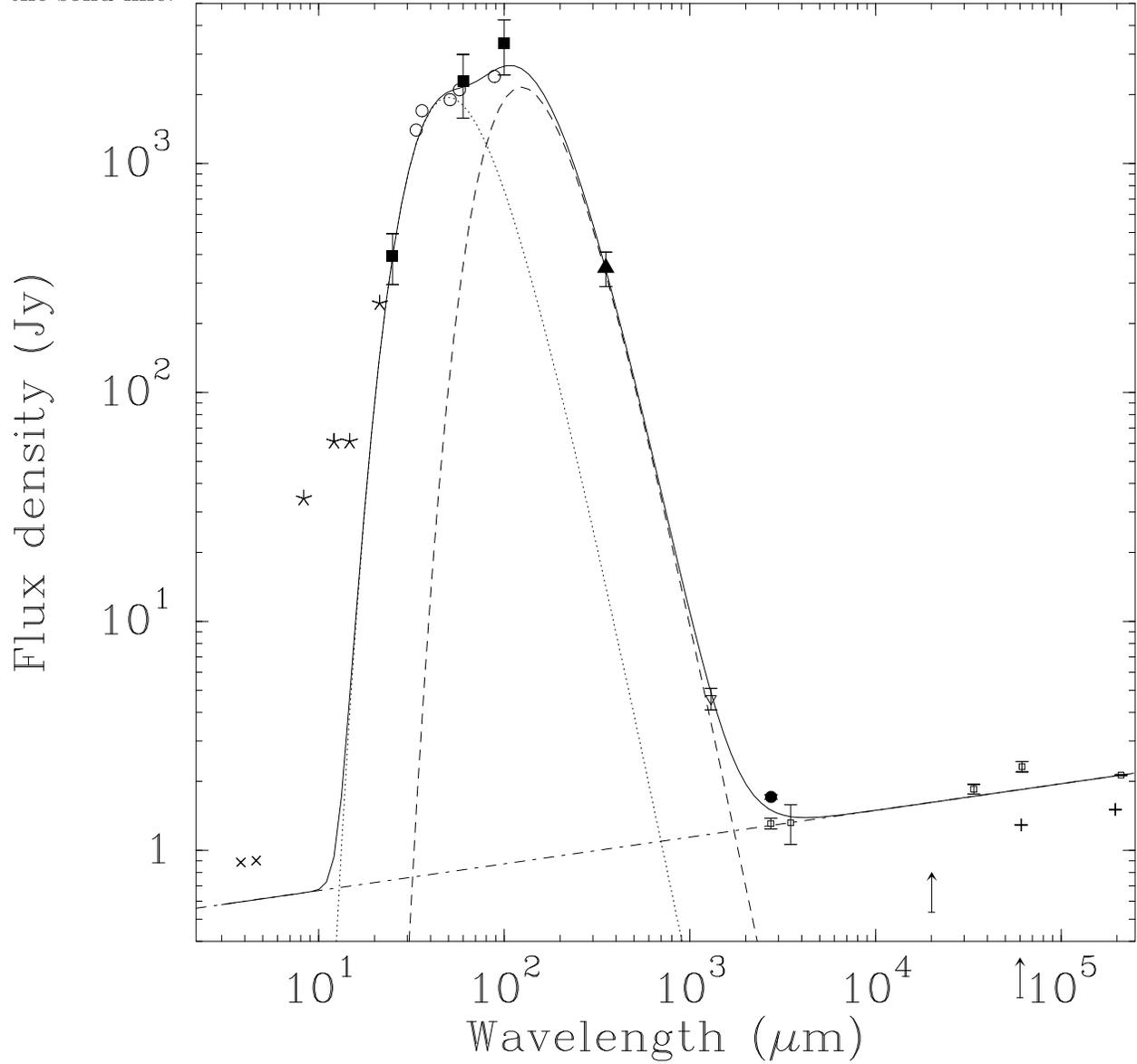]{Global SED of \iras.  The flux density
measurements are summarized in Table~\ref{fluxes} and the components
of the dust and free-free emission models are described in
Table~\ref{sedfits}.  The cold dust component is the dashed line,
the warm dust component is the dotted line, and the free-free
component is the dash-dot line. The sum of all three components
is the solid line.
\label{sed}
}
\epsscale{1.0}
\plotone{f5.eps}
\end{figure}

Using the flux density data from Table~\ref{fluxes}, 
the mid-infrared through radio wavelength spectral energy distribution
(SED) for the entire region is shown in Figure~\ref{sed}.  The flux
density measurements at wavelengths longward of 21~$\mu$m have been
fit with a simple two-temperature modified blackbody dust model plus a
free-free component, summarized in Table~\ref{sedfits}.  The flux
density measurement at 1.3mm from the literature \citep{Chini86}
should be considered a lower limit as it was obtained with a single
element detector with a $90''$ beam centered on the IRAS position,
which misses most of SMM3 and SMM4.  The temperature of the cold
component of dust (25K) agrees quite well with the kinetic temperature
(25.8K) derived from ammonia (1,1) and (2,2) observations with a
$40''$ beam centered on the \uchii\ position \citep{Churchwell90}. It is
interesting to note that the warm component of dust dominates the
total luminosity of the region, which is in contrast to more isolated
high-mass protostellar objects \citep{Sridharan02} and even many other
\uchii\ regions.  Using the grain emissivity index ($\beta$) along with
the temperature and optical depth derived from the fit, one can
calculate the number of cold and warm grains required to explain the
observed flux density \citep{Lonsdale87,Hildebrand83}.  The
corresponding mass of dust can then be calculated for each clump and
for the extended emission.  As is typical, the cold grains dominate
the mass of dust.  Assuming a gas to dust mass ratio of 100
\citep{Sodroski97}, the total gas mass of each clump is listed in
column six of Table~\ref{sharctable} and the total mass of the region
is $\approx 7400$~\msun.  Using the individual gas masses and source
diameters (from the angular diameter and distance), we compute the
column density of hydrogen ($N_{\rm H} = N_{\rm HI} + 2N_{\rm H_2}$)
toward each clump in column seven of Table~\ref{sharctable}.  Finally,
we have estimated the visual and infrared ($K$~band) extinctions toward
each clump by using the conversion formula of $A_V \sim N_{\rm H}/(2
\times 10^{21})$ derived from observations of the ISM at UV
\citep{Whittet81,Bohlin78} and X-ray wavelengths
\citep{Ryter96,Predehl95}, followed by the relation $A_K = 0.112 A_V$
from \citet{Rieke85}.  The extinction values listed in columns eight
and nine of Table~\ref{sharctable} have been further reduced by a
factor of two to more accurately estimate the extinction toward a
young star at the center of the clump, rather than behind it.


\begin{figure}
\figcaption[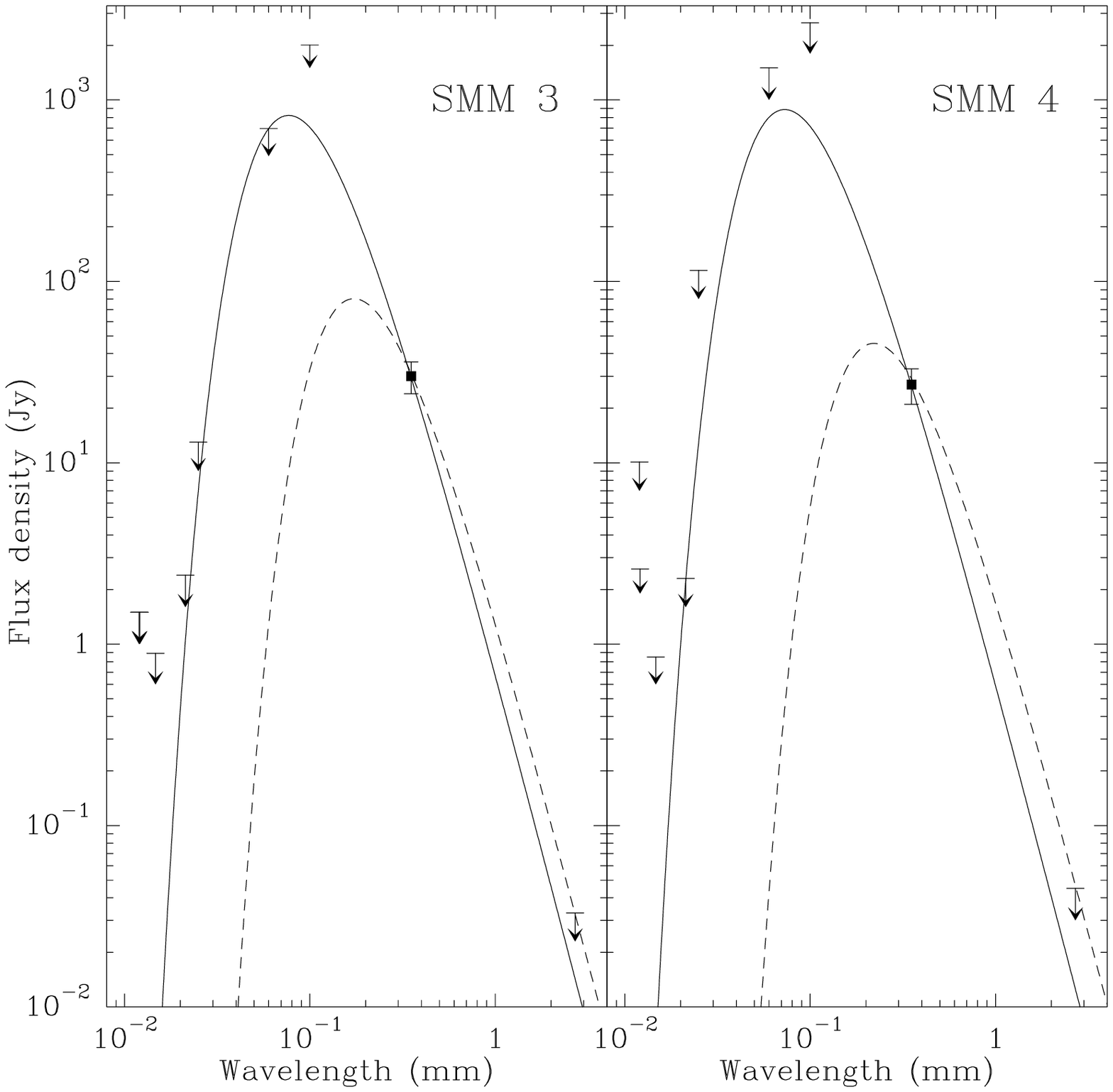]{Individual SEDs of the submillimeter sources
SMM3 and SMM4.  The two longest wavelength flux density measurements
are summarized in Table~\ref{sharctable}.  The infrared upper limits
are from the MSX and IRAS HIRES images.  The solid lines indicate the
warmest dust model consistent with the data, while the dashed lines
indicate the coolest dust model (17-41K for SMM3, 14-41K for SMM4).
\label{smmseds}
}
\epsscale{1.0}
\plotone{f6.eps}
\end{figure}

As the SED model predicts, the free-free emission mechanism still
dominates over the dust emission at frequencies as high as 110~GHz.
In fact, the image at this frequency is nearly identical to the
centimeter images.  The 110~GHz flux densities for the SMM1-5 are
listed in Table~\ref{sharctable}. Nearly all of the 110~GHz flux can
be associated with SMM1, with the rest of the emission sitting just
outside the $22''$ aperture used to define this object in the
$350\mu$m map. By contrast, we have not detected any emission for
SMM2-5.  Each of these upper limits is consistent with an SED
proportional to $\nu^4$, corresponding to dust emission with
$\beta=2$.  For the case of SMM3 and SMM4, they lie sufficiently far
from the main source that useful upper limits can be obtained from
both the IRAS and MSX data which provide a constraint on the
individual properties of these dust cores.  To visualize this
constraint, the spectral energy distributions of SMM3 and SMM4 are
shown in Figure~\ref{smmseds} along with the two most extreme models
consistent with the data.  The corresponding dust temperature and
luminosity upper and lower limits are summarized in
Table~\ref{sed34fits}.  Although the luminosity remains uncertain to
within a factor of 30-60, the lower limits ($\sim1000$\lsun) indicate
that these objects may be powered by individual massive stars or
protostars.

\subsection{\ceo\ (1-0) images}

\begin{figure}
\figcaption[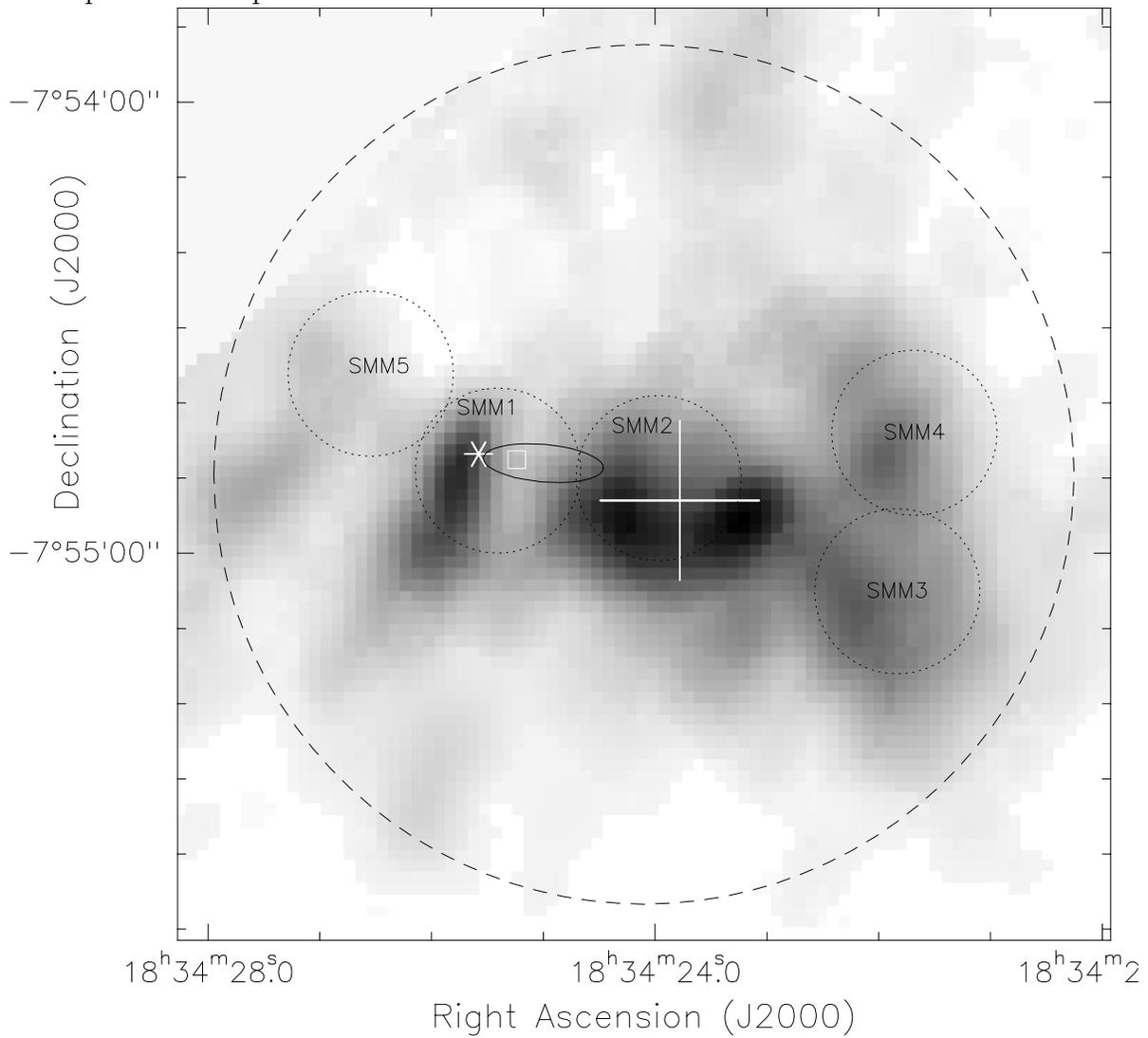]{The integrated emission from
\ceo\ (1-0) is shown in grayscale.  The dashed circle 
marks the BIMA primary beam at
half-maximum response. The dotted circles mark the apertures defining
submillimeter continuum sources SMM1-5. The cross marks the water
maser position uncertainty from \citet{Genzel77}.  The star marks the
peak of the \uchii\ region.  The ellipse is the IRAS position
uncertainty and the square contains the MSX point source position.
\label{c18omom0} 
}
\epsscale{1.0}
\plotone{f7.eps}
\end{figure}

The integrated \ceo\ (1-0) line emission (75-85 km~s$^{-1}$) is shown
as grayscale in Figure~\ref{c18omom0}.  The dotted circles denote the
positions of the submillimeter continuum clumps from
Figure~\ref{sharc}.  The \ceo\ emission has been clipped (set to zero)
at all points below $2.5\sigma$ (0.2 Jy~beam$^{-1}$).  There are five
major peaks of emission, four of which agree roughly with the
submillimeter continuum sources SMM1-4.  The strongest component peaks
very close (offset: $\Delta\alpha,\Delta\delta = +3.0'',-2.6''$) to a
small cluster of stars identified by \citet{Hanson02} that are
associated with the \uchii\ emission and SMM1.  Assuming optically-thin
line emission with $T=25$K, we have computed the total column density
of \ceo\, listed in column 5 of Table~\ref{c18oclumps}.  These values
have been converted to visual extinction $A_V$ using the relationship
of \citet{Hayakawa99} for the Chamaeleon I dark cloud: $N({\rm
C}^{18}{\rm O})({\rm cm}^{-2}) = 3.5 \times 10^{14} A_V - 5.7 \times
10^{14}$.  Next, the values of $A_V$ have been converted $A_K$ (see
section \ref{sedsection}) and these are listed in column 6 of
Table~\ref{c18oclumps}.  Assuming relative abundances of $N_{\rm
H}:N_{\rm CO} = 10^4$ and $N_{\rm CO}:N_{{\rm C}^{18}{\rm O}} = 490$,
the gas mass of each clump has been computed and listed in
column 7 of Table~\ref{c18oclumps}.  
Likewise, the mass of gas associated with each of the continuum
sources SMM1-5 is listed in Table~\ref{smmc18oclumps}. The total 
gas mass (including the
extended emission and all the clumps) is $7300 M_\odot$.  Although
this value is in good agreement with the mass derived independently
from the total 350$\mu$m dust emission, the fraction of mass in the
extended emission (outside of SMM1-5) is 70\% in \ceo\ but only 40\%
in dust.

\begin{figure}
\figcaption[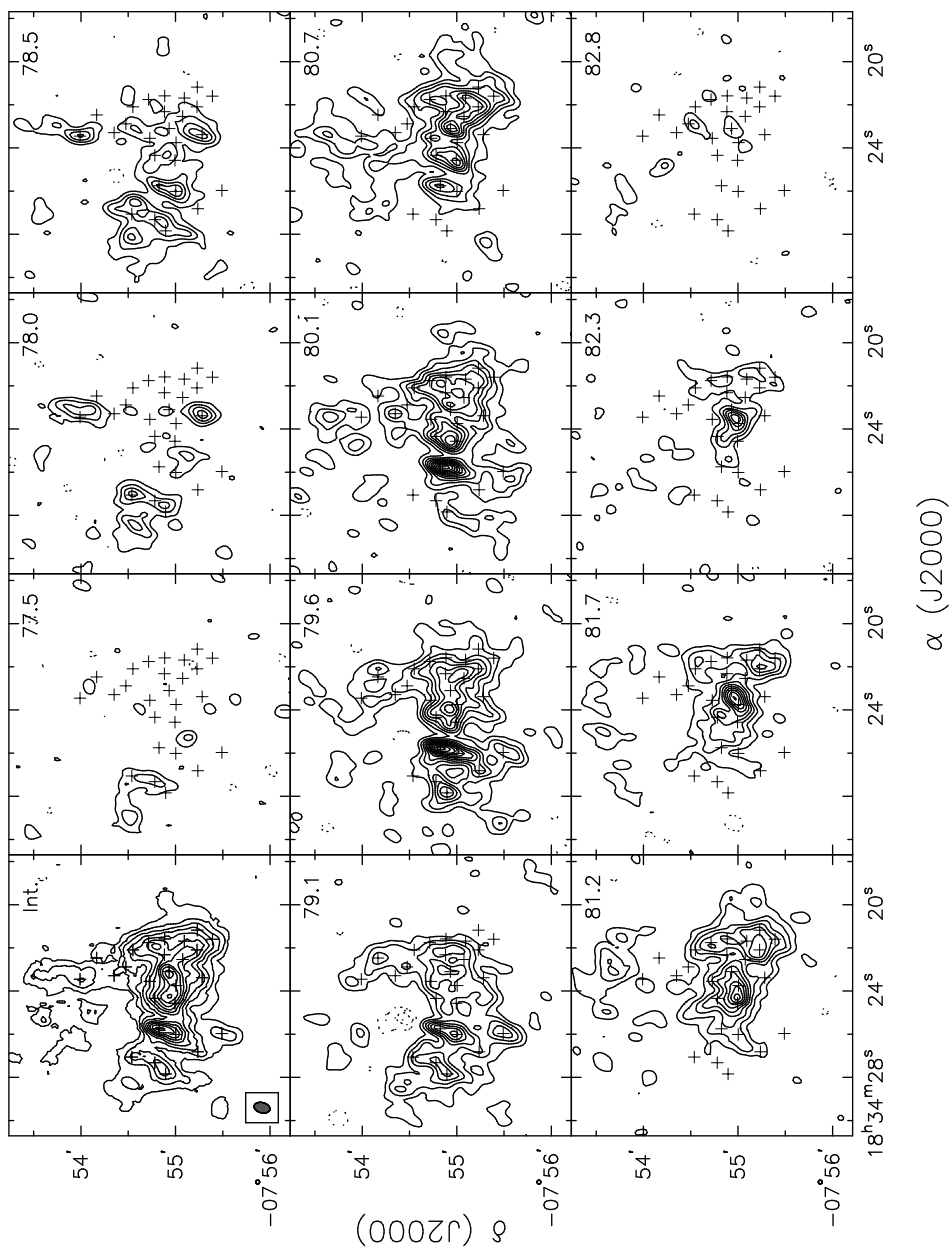]{Channel maps of \ceo\ (1-0).  
The LSR velocity (in km~s$^{-1}$) of the center of the channel 
is given in the upper right corner of each panel. The top left 
panel shows the integrated intensity map, along with the synthesized
beamsize.  The crosses mark the positions of the \ceo\ cores 
identified and listed in Table~\ref{c18oclumps}. Contour levels are 
$+/-0.2 \times (3, 6, 9, 12, 15, 18, 21, 24, 27, 30, 33)$ Jy~beam$^{-1}$.
\label{c18ochanmaps}
}
\epsscale{0.85}
\plotone{f8.eps}
\end{figure}

\begin{figure}
\figcaption[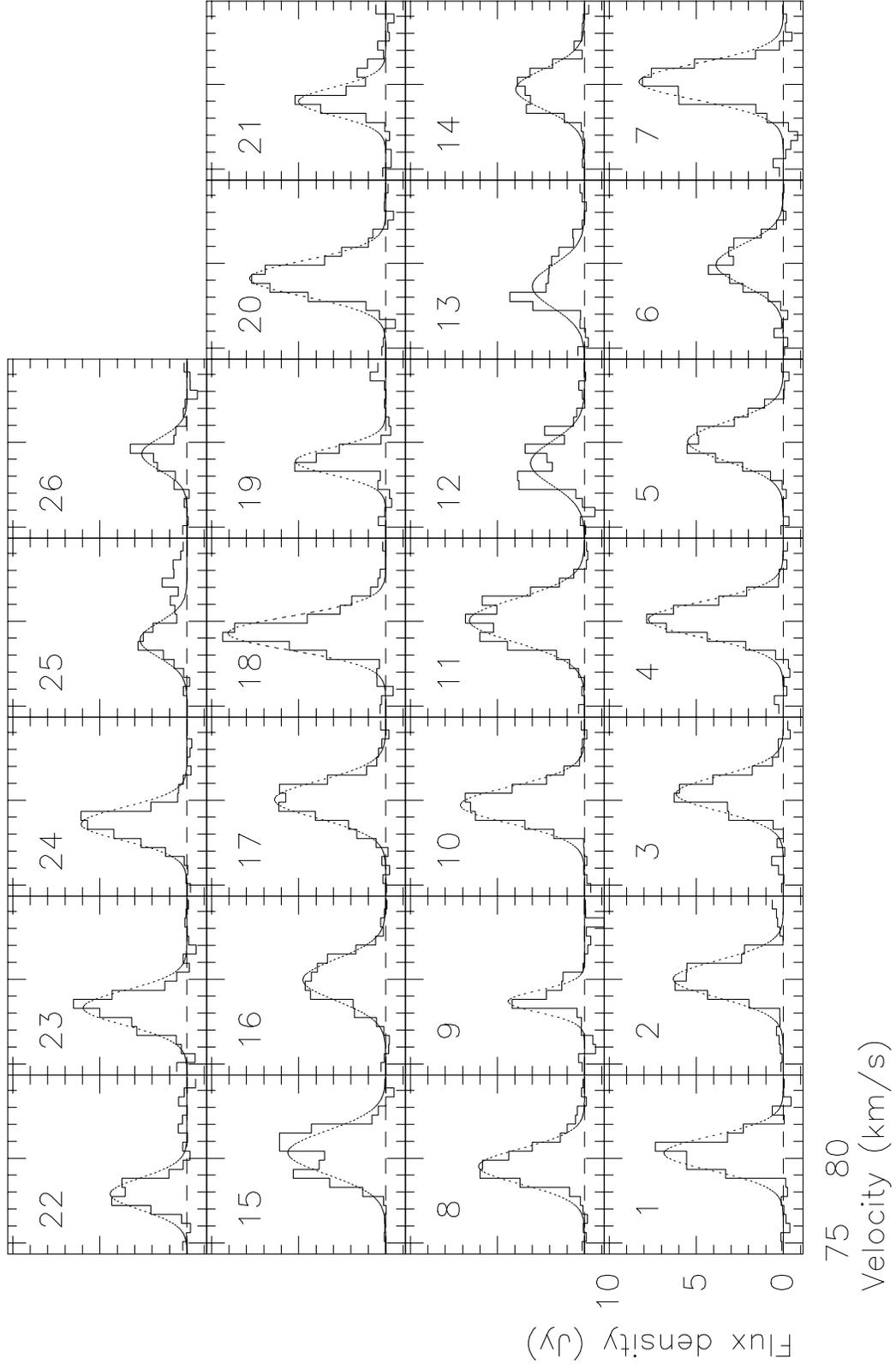]{Spectra from each \ceo\ (1-0) clump
position listed in Table~\ref{c18oclumps}, integrated over a $10''$
diameter aperture. A Gaussian fit to each spectrum is overlaid in
dotted lines.  The fit parameters and the corresponding virial
masses are listed in Table~\ref{c18oclumps}.
\label{c18ospectra}
}
\epsscale{0.84}
\plotone{f9.eps}
\end{figure}

To study the \ceo\ emission in greater detail, channel maps of \ceo\
are shown in Figure~\ref{c18ochanmaps}.  We have analyzed these maps
in two ways: we first inspected the maps visually, then used an
objective computer algorithm.  In the visual method, we manually
identified 26 cores in position-velocity space.  Shown in
Figure~\ref{c18ospectra} is a grid of spectra constructed by
integrating the emission in a $10''$ aperture (0.24 pc) centered at
the position of each core.  The mass contained within these apertures
represents about 40\% of the total gas mass.  

We next attempted to objectively analyze the \ceo\ date cube by
running the ``clumpfind'' program \citep{Williams94}.  This algorithm
contours the data, locates the peaks and follows them to the low
intensity limit without any constraint on the shape of the resulting
clump.  It was designed to operate on large scale maps of GMCs in
which the emission is well separated into distinct clumps.  Our data
do not fit this description, as the cores are embedded in significant
extended emission.  Nevertheless, we proceeded and used the
recommended contour levels by setting both the starting contour and
the contour interval to be twice the RMS of the individual channel
maps (0.6~Jy~km~s$^{-1}$).  The program identified 17 clumps, three of
which were weak and centered slightly outside the primary beam (which
we reject). The largest seven clumps range in mass from
200-1400~\msun, while the smallest seven range from 18-94~\msun.  The
fraction of mass placed into these 14 clumps is 45\% of the total
emission, which is quite similar to our visual technique.  The
emission from the several cores in the western ridge (2,4,5,6,7,8,10)
were merged together by clumpfind into a single large clump in the
late stages of the execution when the lowest contour levels are being
examined.  In a few other cases, two initial clumps merged into one.
This merging effect of the algorithm explains the fewer number (but
larger mass) of clumps found.  The rest of the clumps are in good
general agreement with our visual identification technique.

\begin{figure}
\figcaption[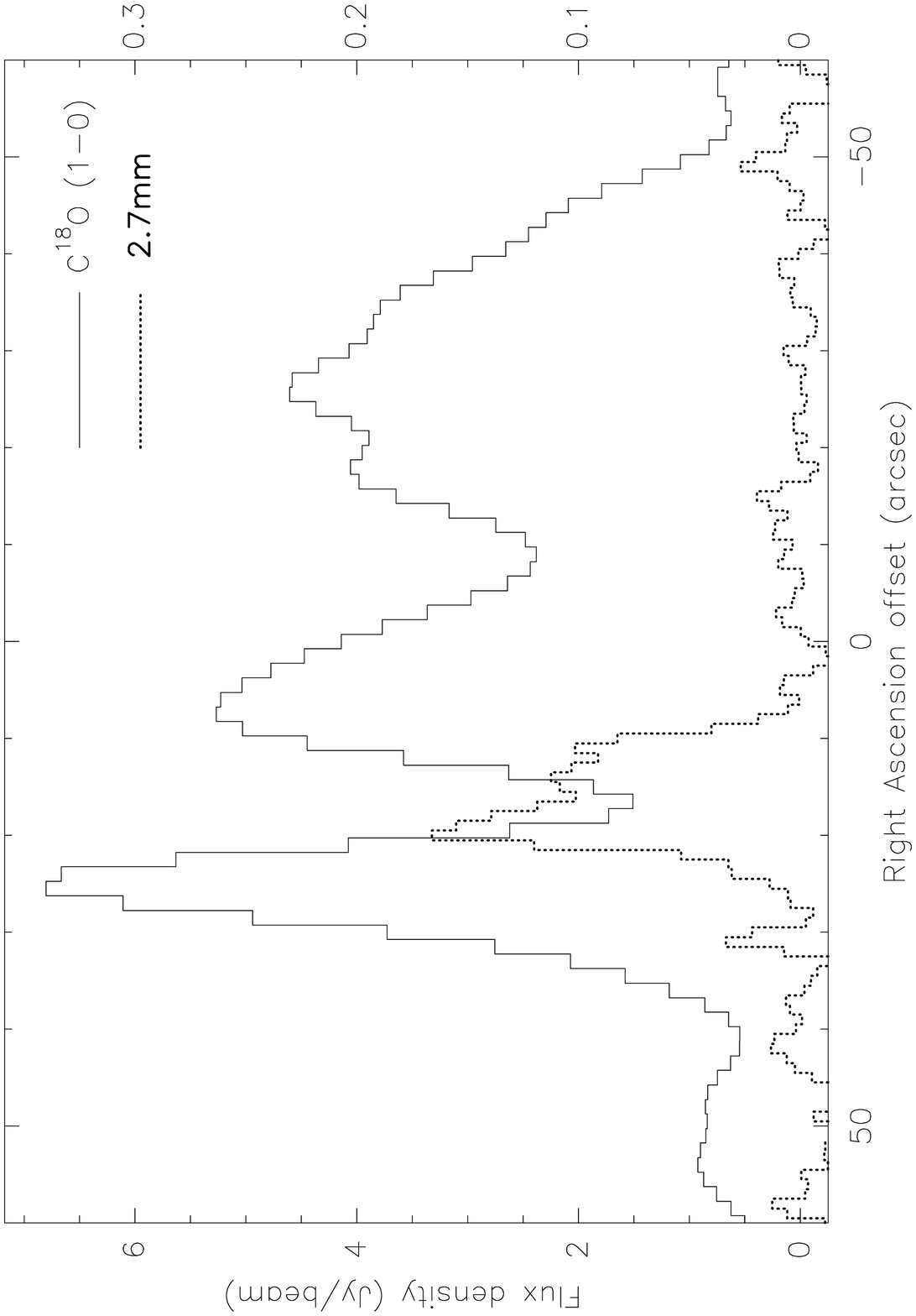]{Solid line: spatial profile of \ceo\
along the position angle $+80^\circ$ passing through clump 18 in the
channel map for velocity 80.1 km~s$^{-1}$. Dotted line: same profile
for the 2.7mm continuum image.  The axis of this strip is indicated in
Figure~\ref{c18omom0_2mass} and Figure~\ref{bimacont}.  The vertical
scale on the left side describes the line data, while the right side 
describes the continuum.
\label{stripchan34}
}
\epsscale{0.80}
\plotone{f10.eps}
\end{figure}

A Gaussian line profile has been fit to each \ceo\ clump, and the
corresponding velocity, amplitude and linewidth is given in
Table~\ref{c18oclumps}.  Using the linewidth ($\delta v$) and aperture
radius ($r$), we compute the virial mass from the formula: $M =
210r$(pc) $\delta v^2$ (km$^2$ s$^{-2}$) \citep{Caselli02}.  In most
cases, and in the overall sum, the virial masses of the clumps are quite
similar to their \ceo-derived masses, suggesting that the clumps are in
hydrostatic equilibrium.  In only three clumps does the \ceo-derived
mass exceed the virial mass by more than 50\%.  The highest excess
(76\%) is seen in clump~18, associated with the \uchii\ region.  Two of
these clumps (17 and 18) lie near the \uchii\ region and also exhibit
the steepest spatial profiles, possibly suggesting an unstable
condition.  A cut along position angle $80^\circ$ in the velocity
channel centered at 80.1 km~s$^{-1}$ is shown in
Figure~\ref{stripchan34}.  The minimum in \ceo\ emission corresponds
to the presence of the free-free continuum emission along the southern
portion of the shell structure seen in Figure~\ref{bimacont}, thus the
steep profile may be due to interaction with the \uchii\ region.  In any
case, considering the uncertainties in the \ceo\ mass calculations,
the good agreement between the \ceo\ mass and the virial mass is
analagous to the results found in a survey of 40 lower-mass \ceo\
clumps in the Taurus complex \citep{Onishi96}.

\subsection{Near-infrared point sources}

\begin{figure}
\figcaption[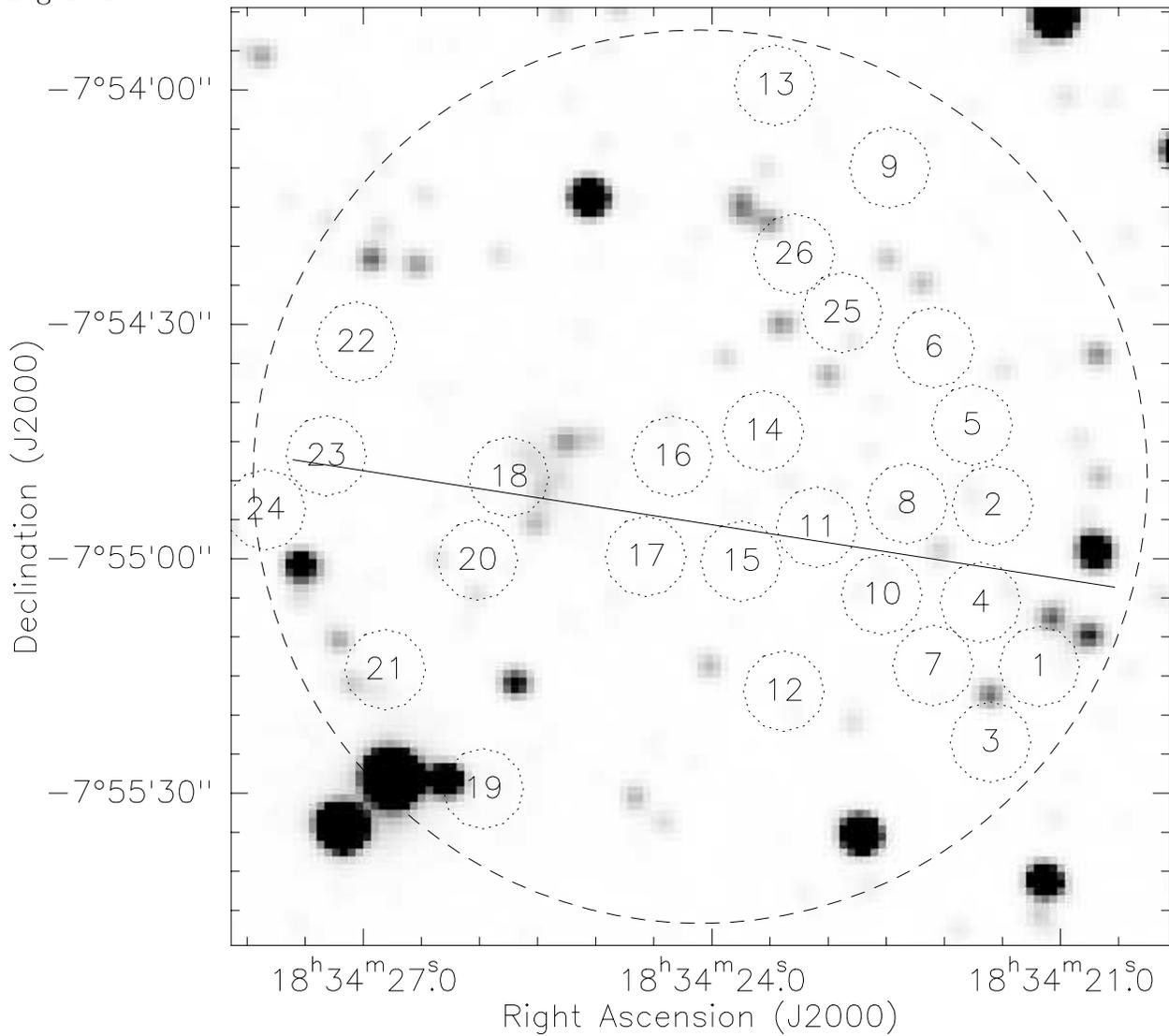]{Near-infrared $K$ band image of
\iras\ from the 2MASS database.  The dotted circles represent the
position of the 26 \ceo\ (1-0) clumps identified in Table~\ref{c18oclumps}
and plotted in Figure~\ref{c18ochanmaps}.  The line indicates the axis
used to produce the emission profiles in Figure~\ref{stripchan34}. 
\label{2massk}
} 
\epsscale{1.0}
\plotone{f11.eps}
\end{figure}

In the 2MASS All-Sky Data Release Point Source Catalog there are 123
objects within a $1'$ radius of the \uchii\ position.  Listed in
Table~\ref{twomass} are the 44 of these stars that are detected in all
three bands.  The $K$ band image is shown in Figure~\ref{2massk} with
the position of the \ceo\ clumps indicated by dotted circles.  In
general, the non-coincidence between the two phenomena is striking.
As listed in column 6 of Table~\ref{c18oclumps}, the extinction at K
band through the \ceo\ clumps ranges from 2.3 to 12.5 magnitudes, with
a median value of 8.1.  The faintest star detected has $K$ magnitude of
14.44 while the brightest upper limit has magnitude 10.11. Thus, even
the brightest $K$ band star observed in the field (star 112 with $M_K =
7.35$) would be undetected if placed behind the typical clump.  This
fact may account for the lack of stars seen toward the \ceo\ clumps.

\begin{figure}
\figcaption[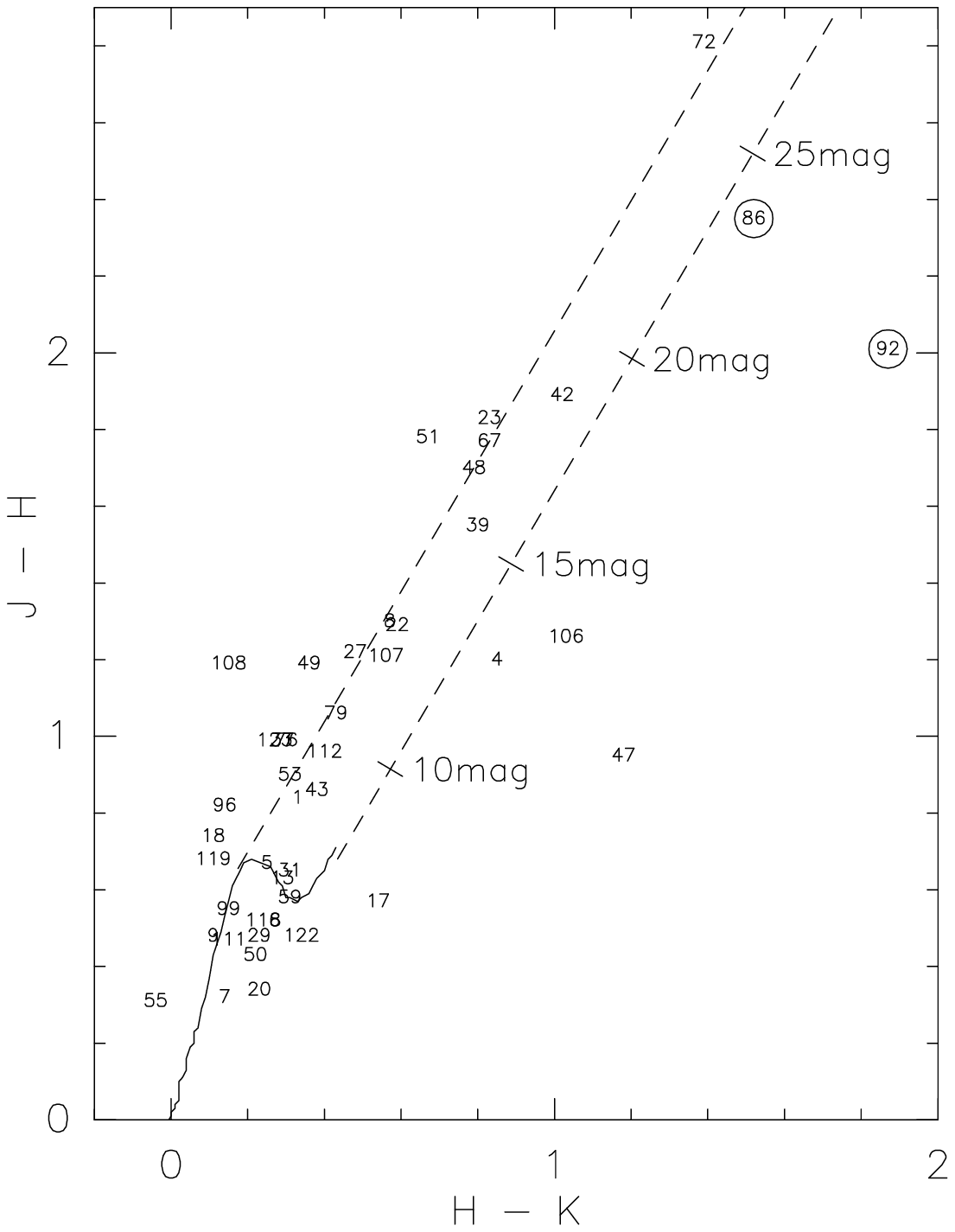]{Color-color diagram ($J-H$ vs. $H-K$) for the stars
detected in all three bands of the 2MASS PSC (numbers correspond to
Table~\ref{twomass}).  The solid line marks the main sequence
\citep{Koornneef83}.  The dashed lines marks the reddening band
\citep{Rieke85} with visual extinction levels marked (for an O7-O9
star). The circled numbers are the two stars associated with the peak
centimeter through submillimeter continuum emission. Of these, star~92
is most closely associated with the \uchii\ region and exhibits an
infrared spectrum consistent with a spectral type of O8.5, while star
86 is classified as O7 \citep{Hanson02}.
\label{colorcolor}
} 
\epsscale{0.80}
\plotone{f12.eps}
\end{figure}

\begin{figure}
\figcaption[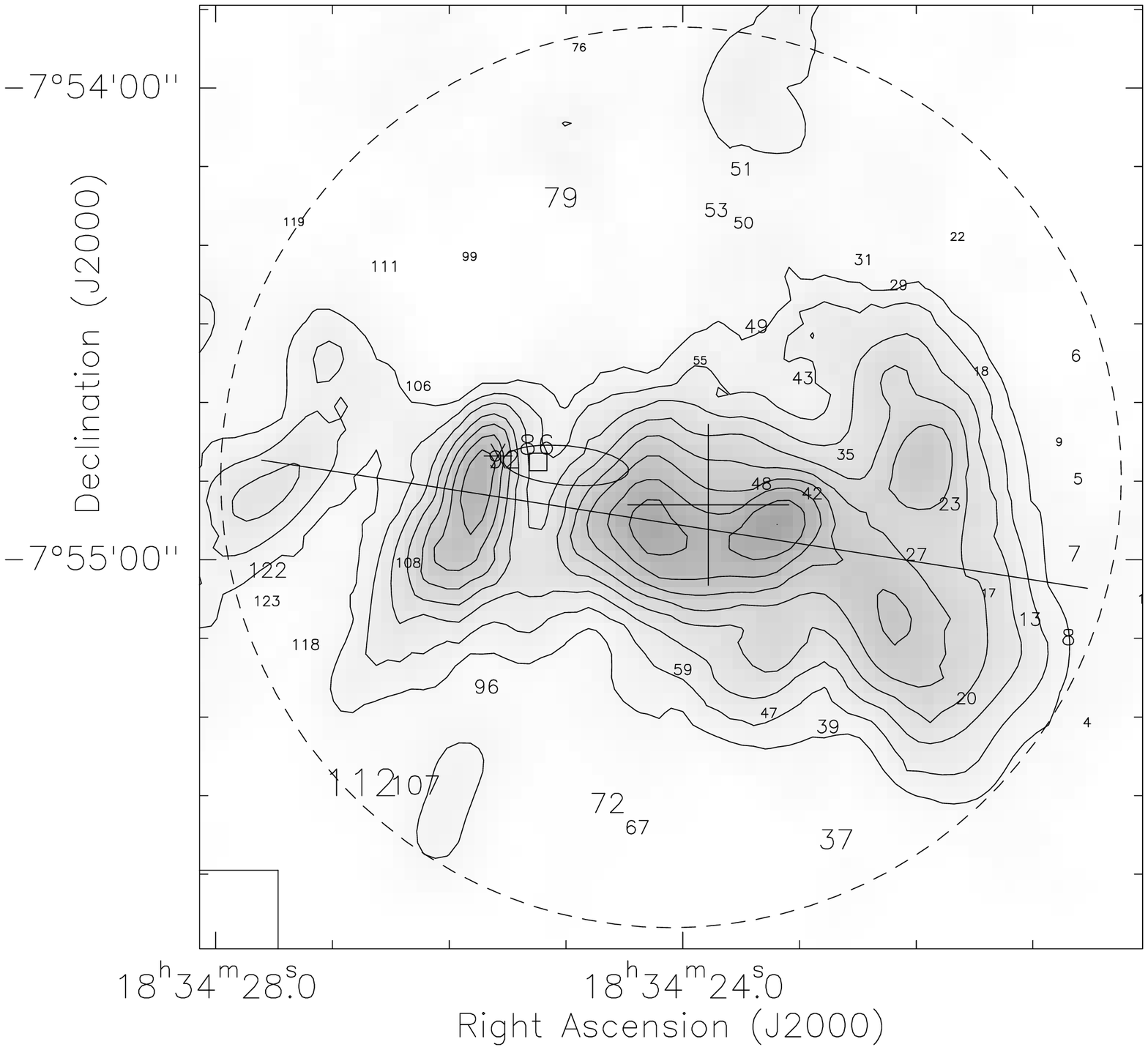]{The integrated emission from \ceo\ (1-0) is shown
in grayscale and contours.
The dashed circle marks the BIMA primary beam at 
half-maximum response. Point sources from the 2MASS 
catalog (detected in all three bands) are indicated by 
their number from Table~\ref{twomass} (with
font size proportional to $K$~band brightness). 
The cross marks the water maser position uncertainty 
from \citet{Genzel77}.  The line indicates
the axis used to produce the emission profiles in
Figure~\ref{stripchan34}. Contour levels are 20\% to 90\% 
of the peak emission (15.8 Jy~beam$^{-1}$).
\label{c18omom0_2mass}
}
\epsscale{1.0}
\plotone{f13.eps}
\end{figure}

A ($J-H$) vs. ($H-K$) color-color diagram of the 2MASS stars is shown
in Figure~\ref{colorcolor}.  The solid line marks the locus of main
sequence stars and the dashed lines denote the reddening vector which
is annotated in magnitudes of visual extinction.  We see that 19 of
the stars exhibit more than 10 magnitudes of visual extinction.
Eleven of these 19 are located within the lowest \ceo\ contours (see
Figure~\ref{c18omom0_2mass}) and are likely to be associated with the
star-forming material of the cluster.  For example, associated with
the \uchii\ region is a small cluster of five stars.  Of these five
stars, star 92 is the object identified as an O8.5 star on the basis
of its infrared spectrum (with weak HeI emission) \citep{Hanson02}.
It lies near the peak of the millimeter continuum map and is one of
the few stars that reside within any of the \ceo\ clumps.  The next
closest star, number 86, lies close to the center of the shell
structure seen in the millimeter continuum.  Due to the presence of
\ion{N}{3} emission, it is classified as an O7 star.  The ratio of
\ion{He}{1} to Br$\gamma$ confirms the level of ionizing flux expected
from such a star.  To within a factor of two, it can account for all
the Lyman continuum flux from the centimeter emission and probably
explains the shell-like symmetry.  Besides stars 86 and 92, three
additional stars (4, 47 and 106) exhibit excess near-infrared emission
(i.e. they lie to the right of the reddening vector in
Figure~\ref{colorcolor}) which may indicate the presence of
circumstellar disks.  Star 106 lies only $4''$ from the center of
SMM5, while star 47 lies at the edge of \ceo\ clump 12.  Star 4 sits
just outside the BIMA primary beam where very little \ceo\ has been
detected.

The brightest star in the field, number 112, has $J$ magnitude 8.7 and
the colors of an M2 star consistent with about 3 magnitudes of visual
extinction.  It could be a foreground giant at 1.8~kpc.  There is no
reference to it in the SIMBAD database.

\section{Discussion} 

\subsection{A protocluster of massive stars?} 

With the exception of the two O-stars associated with SMM1, and star
106 possibly associated with SMM5, none of the near-infrared stars are
associated with the other submillimeter continuum or \ceo\ cores.  The
``starless'' continuum objects SMM2-SMM4 may harbor embedded ZAMS
stars or protostars that are not yet visible in the near-infrared,
while the \ceo\ cores have not yet formed protostars.  If so, they may
resemble the lower mass prestellar cores detected by ISO
\citep{Bacmann00}.  To examine this hypothesis, we can compare the
limiting $K$~band magnitude of the 2MASS image with the expected
brightness of an embedded ZAMS star with total luminosity equal to the
dust luminosity of each core.  It is difficult to estimate the
individual luminosities of these cores due to the limited,
coarse-resolution imaging data available on the mid-infrared side of
their SEDs.  However, using upper limits obtained from the MSX and
IRAS images (as listed in Table~\ref{sed34fits}) the luminosities of
SMM3 and 4 are constrained to be in the range $\approx 1000-40000$
\lsun, depending on the dust temperature.  Assuming the temperature of
25K derived for the region as a whole, yields a typical luminosity of
$2 \times 10^4$ \lsun\, consistent with a B0 star.  A B0 star has
absolute visual magnitude $M_V = -4.1$ \citep{Allen76} and $V-K =
-0.85$ \citep{Koornneef83}, yielding $M_K = -3.25$.  At a distance of
4.9 kpc, this would be reduced to an apparent $K$ magnitude of 8.65.  If
such a star was placed at the center of the SMM4 dust cloud, a $K$
extinction of 6.4 magnitudes (see Table~\ref{sharctable}) would
result, yielding a final $K$ magnitude of $\sim 15$.  By comparison, the
faintest star detected in the 2MASS image has a $K$ magnitude of 14.44.
Thus we cannot rule out the possibility that each submillimeter source
(SMM2-5) may contain a ZAMS or main sequence star rather than a
protostar. Deeper imaging in $K$, $L$ or $M$ band would improve the
constraints.  At present, our best reasonable conclusion is that
SMM2-5 contain some number of young stellar objects or main sequence
stars with luminosities equivalent to at least a B-type star.

With the exception of SMM5, as one moves from east to west across the
region, the general trend is for objects in the cluster to exhibit
fewer signs of compact, energetic phenomena.  SMM1 is associated with
the well-developed \uchii\ region.  The dust core (SMM2) associated with
the water maser probably traces an intermediate stage indicative of
outflow or disk activity from the protostar.  The next two dust cores
(SMM3 and SMM4) exhibit no maser activity or ionized gas.  We note
that the dust-derived masses for these two objects exceed but remain
in reasonable agreement with the \ceo-derived masses (within 27\% and
45\% respectively).  In contrast, the dust-derived mass of the
faintest submillimeter source (SMM5) is a factor of 3 larger than the
\ceo-derived mass.  Unfortunately, the uncertainties in the mass
estimates are too large for us to interpret this difference in
physical terms (such as a depletion of CO, which has been seen in
objects such as B68 by \citet{Bergin02}).  The other molecular cores not
seen in continuum may be the youngest features in the region on their
way to forming stars. Or they could simply be colder, inactive objects
where the accompanying dust emission is below the detection threshold.
Deeper and higher reoslution submillimeter observations are needed to
explore these possibilities.

\subsection{Fragmented structure}

The \ceo\ emission of \iras\ is distributed in clumps aligned roughly
along an east-west ridge.  Evidence of periodic density structure has
been previously observed in \ceo\ in giant molecular clouds,
specifically Orion A \citep{Dutrey91}.  The typical spatial wavelength
they find is 1~parsec, and the fragment masses range from 70-100
$M_\odot$.  More recently, fragmentation has been seen to extend to
even smaller spatial scales in Orion from VLA observations of
\ammonia\ \citep{Wiseman98}.  Similarly, new observations of the
mini-starburst W43 in submillimeter dust continuum reveal about 50
fragments with typical size of 0.25~pc and mass of 300~\msun\
\citep{Motte03}.  In comparison, \ceo\ maps of the Taurus complex
reveal 40 dense cores of a similar typical size as those in W43
(0.23~pc) but with a smaller typical mass of 23 \msun\
\citep{Onishi96}.  In \iras\, the typical spacing we find between the
major \ceo\ cores is roughly $24''$ (0.5 pc), i.e. intermediate
between Orion and W43, while our fragment masses range from 35-187
\msun, i.e. intermediate between W43 and Taurus.  The fraction of
total mass that resides in \ceo\ cores is 40\% which is somewhat
larger than the value of 20\% seen in NH$_3$ cores in W3OH
\citep{Tieftrunk98}, and the 19\% seen in CS cores in Orion B
\citep{Lada91}.  In any case, it is becoming clear that high-mass star
formation regions, like their low-mass counterparts, contain a wealth
of information on the mass spectrum of protostellar fragmentation.
Whether these objects will all form stars remains unclear. A
combination of single-dish and sensitive interferometric studies will
be needed to better quantify the picture, especially down to the low
mass end of the distribution.

\subsection{Future work}

High resolution mid and far infrared imaging is needed to accurately
determine the temperature and size of the individual dust cores
presently identified, and to search for lower mass objects.  Deeper
imaging in the near-infrared is needed to search for additional ZAMS
stars at high extinction levels within the dust and \ceo\ cores.
Also, narrow band imaging in $H_2$ lines and (sub)millimeter
interferometric imaging of SiO transitions may help distinguish which
of the cores show jets and outflows.  Submillimeter interferometry
with higher spectral resolution in other optically-thin tracers less
affected by depletion would be useful to search for evidence of active
infall toward the \ceo\ cores identified in this work.  Finally,
interferometric observations of the 22~GHz or submillimeter water
maser transitions would be quite useful to better localize the maser
activity to SMM2 or one of the \ceo\ cores\footnote{VLA observations
undertaken by the authors on 2004 January 08 (project AH833) in
B-configuration have resolved a pair of 22 GHz water maser spots: one
(at 18:34:23.99, -07:54:48.4) coincident with the submillimeter
continuum source SMM 2 and the other one (at 18:34:24.49, -07:54:47.5)
coincident with the molecular gas clump number 16.}.

\section{Conclusions}

Our high angular resolution observations of the luminous
(log(L/\lsun)=5.2), massive star-forming region \iras\ have revealed a
complex field of objects likely to be in various stages of star
formation.  Of the five submillimeter dust cores, one is associated
with the \uchii\ region G23.955+0.150, and another with a water maser.
The 2.7mm continuum is completely dominated by free-free emission from
the \uchii\ region, with total flux and morphology in agreement with
VLA centimeterwave maps.  For the other four objects, the upper limits
found at 2.7mm and in the MSX mid-infrared band are consistent with
pure optically-thin dust emission at temperatures of 13-40~K and a
dust grain emissivity index $\beta=2$.  Three out of four of these objects
have no associated 2MASS star, and they are each likely to contain at
least one (proto)star of luminosity 1000-40000 \lsun.  In addition, we
have identified two dozen \ceo\ cores in this region which contain
$\approx 40\%$ of the total molecular gas mass ($7300M_\odot$)
present.  Their typical size is 0.25~pc and linewidth is 2-3
km~s$^{-1}$. While the overall extent of the \ceo\ and dust emission
is similar, most of the emission peaks do not correlate well in
detail.  Compared to the dust emission, a greater fraction of the
\ceo\ emission exists in extended features.  At least 11 of the 123
infrared stars identified by 2MASS in this region are likely to be
embedded in the star-forming material, including two O stars powering
the \uchii\ emission. Most of the rest of the reddened stars
anti-correlate with the position of the dust and \ceo\ cores and are
likely visible simply due to the relatively lower extinction.  In
summary, our observations indicate that considerable fragmentation of
the molecular cloud has taken place during the time required for the
\uchii\ region to form and for the O stars to become detectable at
infrared wavelengths.  Additional star formation appears to be ongoing
throughout the region with evidence for up to four B-type (proto)stars
scattered amongst more than two dozen molecular gas cores.

\acknowledgments 

We thank Yu-Nung Su for obtaining the 12-Meter data for us, Robert
Becker for providing freshly-prepared VLA survey images, and Ed
Churchwell for providing valuable comments on the manuscript.  Several
expedient suggestions and corrections to this paper were provided by
the anonymous referee. This research made use of data products from
the Midcourse Space Experiment.  Processing of the data was funded by
the Ballistic Missile Defense Organization with additional support
from NASA Office of Space Science.  This research has also made use of
the NASA/IPAC Infrared Science Archive, which is operated by the Jet
Propulsion Laboratory, California Institute of Technology, under
contract with the National Aeronautics and Space Administration, as
well as the SIMBAD database, operated at CDS, Strasbourg, France

\newpage

\begin{table}[ht]
\small
\caption{Observed emission properties of submillimeter clumps}
\begin{tabular}{crrrrrrrr} 
\tableline 
       & \multicolumn{2}{c}{Coordinates (J2000)} 
       & \multicolumn{2}{c}{Flux density\tablenotemark{a}~(Jy)} 
       & Mass\tablenotemark{d} & log($N_{\rm{H}}$)\tablenotemark{e} 
       & $A_V$\tablenotemark{f}
       & $A_K$\tablenotemark{f}\\
Source & R.A. & Decl. & $350\mu$m 
       & 2.7mm & $M_{\odot}$ & cm$^{-2}$ & mag & mag\\ 
\tableline
SMM1 & 18:34:25.4 & $-07$:54:49 & $63 \pm 6$ & $1.47 \pm 0.03$
     & $1360 \pm 120$ & $23.75\pm 0.04$ & 140 & 15.7\\
SMM2 & 18:34:24.0 & $-07$:54:50 & $67 \pm 6$ & $<0.033$\tablenotemark{b} 
     & $1400\pm 120$ & $23.76\pm 0.04$ & 144 & 16.1\\
SMM3 & 18:34:21.8 & $-07$:55:05 & $30 \pm 6$ & $<0.033$\tablenotemark{b} 
     & $640\pm 120$  & $23.42\pm 0.08$ &  66 &  7.4\\
SMM4 & 18:34:21.7 & $-07$:54:44 & $27 \pm 6$ & $<0.045$\tablenotemark{b} 
     & $560\pm 120$  & $23.36\pm 0.09$ &  57 &  6.4\\
SMM5 & 18:34:26.5 & $-07$:54:36 & $24 \pm 6$ & $<0.033$\tablenotemark{b} 
     & $500\pm 120$  & $23.32\pm 0.09$ &  52 &  5.8 \\ 
Extended &       &     & $140 \pm 20$ & $0.25 \pm 0.03$ 
     & $2900\pm 300$ &\\
\tableline 
Total\tablenotemark{c}&   &     & $350 \pm 60$ & $1.72 \pm 0.06$ 
     & $7400 \pm 900$&\\
\end{tabular} 
\tablenotetext{a}{Within $22''$ aperture, as shown in Figure~\ref{sharc}}
\tablenotetext{b}{Upper limits are $3\sigma$}
\tablenotetext{c}{Total flux including extended emission within dotted 
region of Figure~\ref{sharc}}
\tablenotetext{d}{Assuming T=25K and grain emissivity Q=1.0E-04 at $350\mu$m \citep{Hildebrand83}} 
\tablenotetext{e}{Assuming uniform emission of diameter of $11''$ (0.25~pc)}
\tablenotetext{f}{Extinction to a star at the center of the dust core, i.e. 
half the column density}
\label{sharctable}
\end{table}

\begin{deluxetable}{cccccc}
\tabletypesize{\small}
\tablewidth{0pt}
\tablecaption{Summary of flux density measurements of \iras}
\tablehead{
 \colhead{Frequency (GHz)} & \colhead{Wavelength ($\mu$m)} & 
 \colhead{Flux (Jy)} & \colhead{Aperture/beamsize} & \colhead{Instrument} & 
 \colhead{Reference}
}
\startdata
      & 3.8  & 0.885\tablenotemark{a} & $15''$ & CFHT & 1\\
      & 4.6  & 0.902\tablenotemark{a} & $15''$ & CFHT & 1\\
      & 8.3  & $34.4 \pm 1.7$         & $21''$ & MSX  & this work\\
      & 12   & $66.3 \pm 6.6$         & $62'' \times 27''$ & IRAS & 2\\
      & 12.1 & $61 \pm 2$             & $22''$ & MSX  & this work\\
      & 14.7 & $61 \pm 2$             & $22''$ & MSX  & this work\\
      & 21.3 & $245 \pm 15$           & $23''$ & MSX  & this work\\
      & 25   & $395 \pm 99$           & $84'' \times 29''$ & IRAS & 2\\
      & 33.5 & $1400 \pm 100$         & $44''$ & KAO & 3\\
      & 36   & $1700 \pm 100$         & $44''$ & KAO & 3\\
      & 51   & $1900 \pm 100$         & $44''$ & KAO & 3\\
      & 57.3 & $2100 \pm 100$         & $44''$ & KAO & 3\\
      & 60   & $2285 \pm 708$         & $144'' \times 62''$ & IRAS & 2\\
      & 88.4 & $2400 \pm 100$         & $44''$ & KAO & 3\\
      & 100  & $3339 \pm 902$         & $152'' \times 137''$ & IRAS & 2\\
857   & 352  & $320 \pm 64$           &        & CSO/SHARC & 4\\
857   & 352  & $350 \pm 60$           & $98'' \times 44''$ & CSO/SHARC & this work\\
230   & 1.3 mm   & $4.6 \pm 0.5$      & $90''$    & IRTF & 2\\
110   & 2.725 mm & $1.31 \pm 0.07$    & $54''$    & NRAO 12m & 5\\
110   & 2.735 mm & $1.72 \pm .04$     &           & BIMA & this work\\
 86   & 3.486 mm & $1.32 \pm 0.20$    & $78''$    & NRAO 11m & 6\\
14.94 & 2.007 cm & 0.536   \tablenotemark{b} &    & VLA-B & 7\\
14.8  & 2.0 cm   & $1.56 \pm 0.08$    & $60''$    & Effelsberg 100m & 6\\
10.3  & 2.91 cm  & $2.11 \pm 0.04$    & $160''$   & NRO-45m & 8\\
8.875 & 3.378 cm & $1.85 \pm 0.09$    & $84''$    & Effelsberg 100m & 6\\
4.9   & 6.1 cm   & $1.290 \pm 0.003$  & $9''\times 4''$     & VLA-B & 9\\
4.875 & 6.15 cm  & $2.32 \pm 0.12$    & $2.6'$    & Effelsberg 100m & 6\\
4.875 & 6.15 cm  & $2.50 \pm 0.13$    & $2.6'$    & Effelsberg 100m & 10\\
4.860 & 6.17 cm  & 0.227 \tablenotemark{b} &      & VLA-A/B & 7\\
1.527 & 19.6 cm  & $1.508 \pm 0.005$  &           & VLA-B   & 9\\
1.527 & 19.6 cm  & $1.608 \pm 0.040$  & $7''\times4''$ & VLA-B   & 11\\
1.425 & 21.0 cm  & $2.13\pm 0.01$     &           & VLA-DnC & 12\\
\enddata
\tablenotetext{a}{aperture did not cover the \uchii\ position}
\tablenotetext{b}{measurements suffer from missing flux}
References: (1) \citet{Chini87}; (2) \citet{Chini86}; (3) \citet{Simpson95};
(4) \citet{Mueller02}; (5) \citet{Wood88}; (6) \citet{Wink82};
(7) \citet{Wood89}; (8) \citet{Handa87}; (9) \citet{Becker94};
(10) \citet{Altenhoff79}; (11) \citet{Garwood88}; (12) \citet{Kim01}.
\label{fluxes}
\end{deluxetable}

\begin{table}[ht]
\small
\caption{Parameters of the global SED model}
\begin{tabular}{crrrcc} 
\tableline 
Component & T (K) & $\beta$ & $\tau_{\rm 125{\mu}m}$ & L (\lsun) & M (\msun)\\
\tableline
cold dust & 25 & 2.0 & 0.38  &   50,000 & 7200 \\
warm dust & 63 & 2.0 & 0.062 &  110,000 & 70 \\
\tableline
total dust &   &     &       &  160,000 & 7300\\
\tableline
free-free & \multicolumn{5}{l}{$F_\nu(Jy)=2.3\nu_{\rm GHz}^{-0.117}$}\\
\tableline  
\end{tabular} 
\label{sedfits}
\end{table}

\begin{table}[ht]
\small
\caption{Range of parameters for the SED models of SMM3 and SMM4}
\begin{tabular}{ccrrrcc}
\tableline 
Component & Infrared limit & T (K) & $\beta^a$ & $\tau_{\rm 125{\mu}m}$ & L (\lsun) & M (\msun)\\
\tableline
SMM3 & $<696$Jy at $60\mu$m & 17-41 & 2.0 & 0.30-0.049 & 1400-39000 & 270-1600\\
SMM4 & $<2.3$Jy at $21.3\mu$m & 14-41 & 2.0 & 0.48-0.044 & 630-38000 & 240-2600\\
\tableline
$^a$Value fixed in fit\\
\end{tabular} 
\label{sed34fits}
\end{table}

\begin{deluxetable}{crrccccccccc}
\tabletypesize{\scriptsize}
\tablewidth{0pt}
\tablecaption{Position and mass of \ceo\ (1-0) clumps\label{c18oclumps}}
\tablehead{
 \colhead{ } & \colhead{R.A.} & \colhead{Dec} & \colhead{Emission} & 
 \colhead{log(\nceo)} 
 & \colhead{$A_K$\tablenotemark{a}}  
 & \colhead{$M_{H_2}$\tablenotemark{b}}  
 & \colhead{Velocity} & \colhead{Peak flux} &  
 \colhead{Linewidth} & \colhead{$M_{\rm virial}$}\\
 \colhead{\#} & \colhead{J2000} & \colhead{J2000} & \colhead{Jy km~s$^{-1}$}  
 & \colhead{cm$^{-2}$} & \colhead{mag.}  &
 \colhead{M$_\odot$} & \colhead{km~s$^{-1}$} & \colhead{Jy beam$^{-1}$} & 
 \colhead{km~s$^{-1}$} & \colhead{M$_\odot$}\\
}
\startdata
1 &  18:34:21.183   & -07:55:13.78   & 14.3 & 16.39 & 8.1 & 119 & $80.3\pm 0.1$ & $6.9\pm 0.6$ & $2.1\pm 0.2$ & 110\\
2 &  18:34:21.583   &  -07:54:53.17  & 16.7 & 16.46 & 9.5 & 139 & $80.0\pm 0.1$ & $6.3\pm 0.6$ & $2.3\pm 0.2$ & 120\\
3 &  18:34:21.602   & -07:55:23.45   & 14.3 & 16.39 & 8.1 & 120 & $80.4\pm 0.2$ & $6.2\pm 0.6$ & $2.1\pm 0.2$ & 110\\
4 &  18:34:21.689   &  -07:55:05.58  & 17.9 & 16.49 &10.2 & 150 & $80.1\pm 0.1$ & $7.8\pm 0.6$ & $2.2\pm 0.2$ & 115\\
5 &  18:34:21.760   &  -07:54:42.88  & 14.8 & 16.40 & 8.3 & 123 & $80.0\pm 0.1$ & $5.5\pm 0.5$ & $2.6\pm 0.3$ & 136\\
6 &  18:34:22.090   &  -07:54:32.96  & 11.3 & 16.29 & 6.5 &  94 & $79.9\pm 0.2$ & $3.9\pm 0.5$ & $2.8\pm 0.4$ & 147\\
7 &  18:34:22.097   &  -07:55:13.66  & 18.9 & 16.51 &10.6 & 158 & $80.2\pm 0.1$ & $8.3\pm 0.6$ & $2.3\pm 0.2$ & 120\\
8 &  18:34:22.321   &  -07:54:52.93  & 14.7 & 16.40 & 8.3 & 123 & $79.5\pm 0.1$ & $6.1\pm 0.6$ & $2.2\pm 0.2$ & 115\\
9 &  18:34:22.470   &  -07:54:09.96  & 3.8  & 15.82 & 2.3 &  32 & $78.7\pm 1.0$ & $4.4\pm 0.9$ & $0.8\pm 0.2$ & 42\\
10 &  18:34:22.539  &  -07:55:04.57  & 17.3 & 16.48 & 9.9 & 145 & $79.7\pm 0.1$ & $7.1\pm 0.6$ & $2.3\pm 0.2$ & 120\\
11 &  18:34:23.098  &  -07:54:55.95  & 22.1 & 16.58 &12.5 & 185 & $80.0\pm 0.1$ & $6.6\pm 0.5$ & $3.2\pm 0.2$ & 168\\
12 &  18:34:23.381  &  -07:55:16.93  & 10.7 & 16.27 & 6.2 &  90 & $78.8\pm 0.3$ & $3.1\pm 0.5$ & $3.4\pm 0.5$ & 154\\
13 &  18:34:23.457  &  -07:53:59.41  & 10.3 & 16.25 & 5.9 &  86 & $78.8\pm 0.3$ & $3.0\pm 0.5$ & $3.2\pm 0.6$ & 168\\
14 &  18:34:23.558  &  -07:54:43.66  & 11.3 & 16.29 & 6.5 &  94 & $79.7\pm 0.2$ & $3.9\pm 0.5$ & $2.7\pm 0.4$ & 122\\
15 &  18:34:23.750  &  -07:55:00.31  & 18.5 & 16.51 &10.6 & 155 & $80.3\pm 0.2$ & $5.6\pm 0.5$ & $3.3\pm 0.3$ & 161\\
16 &  18:34:24.341  &  -07:54:46.89  & 14.7 & 16.40 & 8.3 & 123 & $80.0\pm 0.2$ & $4.8\pm 0.5$ & $2.9\pm 0.3$ & 131\\
17 &  18:34:24.570  &  -07:54:59.70  & 17.1 & 16.47 & 9.7 & 143 & $80.0\pm 0.1$ & $6.4\pm 0.5$ & $2.5\pm 0.2$ & 113\\
18 &  18:34:25.749  &  -07:54:49.53  & 21.0 & 16.56 &11.9 & 176 & $79.3\pm 0.1$ & $9.1\pm 0.6$ & $2.0\pm 0.2$ & 100\\
19 &  18:34:25.968  &  -07:55:29.42  & 7.4  & 16.11 & 4.3 &  62 & $78.9\pm 0.1$ & $5.2\pm 0.8$ & $1.2\pm 0.2$ & 36\\
20 &  18:34:26.017  &  -07:55:00.15  & 18.3 & 16.50 &10.4 & 153 & $79.1\pm 0.1$ & $7.8\pm 0.6$ & $2.2\pm 0.2$ & 115\\
21 &  18:34:26.812  &  -07:55:14.18  & 8.5  & 16.17 & 5.0 &  71 & $79.0\pm 0.1$ & $5.0\pm 0.7$ & $1.4\pm 0.3$ & 49\\
22 &  18:34:27.061  &  -07:54:32.24  & 8.6  & 16.17 & 5.0 &  72 & $77.9\pm0.2$  & $4.4\pm 0.6$ & $1.8\pm 0.3$ & 81\\
23 &  18:34:27.319  &  -07:54:46.83  & 13.0 & 16.35 & 7.4 & 108 & $78.3\pm0.1$  & $6.0\pm 0.6$ & $2.1\pm 0.2$ & 110\\
24 &  18:34:27.843  &  -07:54:53.72  & 12.0 & 16.31 & 6.8 & 100 & $78.6\pm 0.1$  & $6.1\pm 0.6$ & $1.9\pm 0.2$ & 81\\ 
25 &  18:34:22.880  &  -07:54:28.50  &  8.5 & 15.25 & 5.0 & 71  & $78.9\pm 0.3$ & $2.7\pm 0.6$ & $2.1\pm 0.6$ & 110\\
26 &  18:34:23.292  &  -07:54:21.00  &  5.0 & 15.02 & 3.0 & 42  & $79.3\pm 0.3$ & $2.6\pm 0.6$ & $1.8\pm 0.6$ &  81\\

\tableline
\multicolumn{3}{l}{Total clump component} & 352 & & & 2930 & & & & 2915\\
\multicolumn{3}{l}{Extended component}    & 527 & & & 4410 & & \\  
\multicolumn{3}{l}{Total emission}        & 879 & & & 7340\\  
\enddata
\tablenotetext{a}{$A_V$ computed from \nceo\ using formula from \citet{Hayakawa99}, 
and converted to $A_K$ using \citet{Rieke85}.}
\tablenotetext{b}{Mass computed using formula from \citet{Scoville86}
assuming optically-thin gas at 25K and [$^{12}$CO/\ceo]=490 and [H$_2$/CO]=10$^4$, yielding a conversion factor of 8.35\msun\ (Jy~km~s$^{-1}$)$^{-1}$.} 
\end{deluxetable}

\begin{table}[ht]
\small
\caption{Mass of \ceo\ (1-0) emission associated with SMM objects}
\begin{tabular}{crrrcc} 
\tableline 
   & Emission\tablenotemark{a} & Mass\tablenotemark{b}\\
\# & Jy km~s$^{-1}$  & M$_\odot$\\
\tableline 
SMM1 & 53.5    &  450 \\
SMM2 & 72.7    &  610\\
SMM3 & 60.1    &  500\\
SMM4 & 45.9    &  380\\
SMM5 & 20.3    &  170\\
Extended & 620 & 5200\\
\tableline
Total & 870    & 7300\\
\tableline
\tablenotetext{a}{Not corrected for primary beam attenuation}
\tablenotetext{b}{Mass computed using formula from \citet{Scoville86}
assuming optically-thin gas at 25K and [$^{12}$CO/\ceo]=490 and 
[H$_2$/CO]=10$^4$.} 
\end{tabular} 
\label{smmc18oclumps}
\end{table}

\begin{deluxetable}{rc}
\tabletypesize{\small}
\tablewidth{0pt}
\tablecaption{Stars detected in all three 2MASS bands\label{twomass}}
\tablehead{ \colhead{\#} & \colhead{2MASS PSC}}
\startdata
1 & 18342011-0755050 \\
4\tablenotemark{a} & 18342057-0755207 \\
5 & 18342064-0754497 \\
6 & 18342066-0754341 \\
7 & 18342067-0754593 \\
8\tablenotemark{a} & 18342072-0755099 \\
9 & 18342080-0754450 \\
13 & 18342105-0755076 \\
18 & 18342147-0754360 \\
20 & 18342159-0755177 \\
22\tablenotemark{a} & 18342166-0754189 \\
23\tablenotemark{a} & 18342173-0754530 \\
27\tablenotemark{a} & 18342201-0754594 \\
29 & 18342216-0754251 \\
31 & 18342246-0754219 \\
37 & 18342269-0755357 \\
39\tablenotemark{a} & 18342276-0755213 \\
42\tablenotemark{a} & 18342290-0754516 \\
43 & 18342298-0754369 \\
47\tablenotemark{a} & 18342326-0755195 \\
48\tablenotemark{a} & 18342333-0754504 \\
49\tablenotemark{a} & 18342337-0754304 \\
50 & 18342348-0754172 \\
51\tablenotemark{a} & 18342349-0754103 \\
53 & 18342371-0754156 \\
55 & 18342385-0754347 \\
59 & 18342399-0755140 \\
67\tablenotemark{a} & 18342438-0755341 \\ 
72\tablenotemark{a} & 18342463-0755311 \\
76 & 18342487-0753549 \\
79\tablenotemark{a} & 18342503-0754140 \\
86\tablenotemark{a} & 18342523-0754455 \\
92\tablenotemark{a} & 18342551-0754473 \\
96 & 18342566-0755162 \\
99 & 18342580-0754214 \\
106\tablenotemark{a} & 18342624-0754379 \\
107\tablenotemark{a} & 18342627-0755288 \\
108\tablenotemark{a} & 18342633-0755004 \\
111 & 18342652-0754227 \\
112 & 18342673-0755285 \\
118 & 18342720-0755108 \\
119 & 18342730-0754170 \\
122 & 18342752-0755014 \\
123 & 18342753-0755053 \\
\enddata
\tablenotetext{a}{Stars with $A_V > 10$, according to the color-color 
diagram shown in Figure~\ref{colorcolor}}
\end{deluxetable}

\end{document}